\documentclass[aps,pra,superscriptaddress,nofootinbib,showpacs,floatfix,twocolumn]{revtex4-1}

\usepackage{graphicx}
\usepackage{epstopdf}
\usepackage{amsmath}
\usepackage{amssymb}
\usepackage{hyperref}
\usepackage{bm}
\usepackage{subfigure}

\hypersetup{
    colorlinks=true,       
    linkcolor=cyan,          
    citecolor=magenta,        
    filecolor=magenta,      
    urlcolor=cyan,           
    runcolor=cyan
}
\usepackage[pdftex]{color}

\newcommand{\beq}{\begin{equation}}
\newcommand{\eeq}{\end{equation}}
\newcommand{\beqnn}{\begin{equation*}}
\newcommand{\eeqnn}{\end{equation*}}
\newcommand{\bea}{\begin{eqnarray}}
\newcommand{\eea}{\end{eqnarray}}
\newcommand{\beann}{\begin{eqnarray*}}
\newcommand{\eeann}{\end{eqnarray*}}
\newcommand{\bes} {\begin{subequations}}
\newcommand{\ees} {\end{subequations}}

\newcommand{\braket}[2]{\langle #1 | #2\rangle}
\newcommand{\ket}[1]{ | #1\rangle}
\newcommand{\bra}[1]{\langle #1 | }

\newcommand{\Tr}{\mathrm{Tr}}
\newcommand{\eps}{\varepsilon}

\newcommand{\ignore}[1]{}

\begin{document}
\title{Reexamination of the evidence for entanglement in a quantum annealer}

\author{Tameem Albash}
\email{albash@usc.edu}
\affiliation{Department of Physics and Astronomy, University of Southern California, Los Angeles, California 90089, USA}
\affiliation{Information Sciences Institute, University of Southern California, Marina del Rey, CA 90292}
\affiliation{Center for Quantum Information Science \& Technology, University of Southern California, Los Angeles, California 90089, USA}

\author{Itay Hen}
\affiliation{Information Sciences Institute, University of Southern California, Marina del Rey, CA 90292}
\affiliation{Center for Quantum Information Science \& Technology, University of Southern California, Los Angeles, California 90089, USA}

\author{Federico M. Spedalieri}
\affiliation{Information Sciences Institute, University of Southern California, Marina del Rey, CA 90292}
\affiliation{Center for Quantum Information Science \& Technology, University of Southern California, Los Angeles, California 90089, USA}
\affiliation{Department of Electrical Engineering, University of Southern California, Los Angeles, California 90089, USA}

\author{Daniel A. Lidar}
\affiliation{Department of Physics and Astronomy, University of Southern California, Los Angeles, California 90089, USA}
\affiliation{Center for Quantum Information Science \& Technology, University of Southern California, Los Angeles, California 90089, USA}
\affiliation{Department of Electrical Engineering, University of Southern California, Los Angeles, California 90089, USA}
\affiliation{Department of Chemistry, University of Southern California, Los Angeles, California 90089, USA}

\begin{abstract}
A recent experiment [Lanting \textit{et al.}, PRX, (2014)] claimed to provide evidence of up to $8$-qubit entanglement in a D-Wave quantum annealing device. However, entanglement was measured using qubit tunneling spectroscopy, a technique that provides indirect access to the state of the system at intermediate times during the anneal by performing measurements at the end of the anneal with a probe qubit.  
In addition, an underlying assumption was that the quantum transverse-field Ising Hamiltonian, whose ground states are already highly entangled,  is an appropriate model of the device, and not some other (possibly classical) model. This begs the question of whether alternative, classical or semiclassical models would be equally effective at predicting the observed spectrum and thermal state populations. To check this, we consider a recently proposed classical rotor model with classical Monte Carlo updates, which has been successfully employed in describing features of earlier experiments involving the device. We also consider simulated quantum annealing with quantum Monte Carlo updates, an algorithm that samples from the instantaneous Gibbs state of the device Hamiltonian. Finally, we use the quantum adiabatic master equation, which cannot be efficiently simulated classically, and which has previously been used to successfully capture the open system quantum dynamics of the device.  
We find that only the master equation is able to reproduce the features of the tunneling spectroscopy experiment, while both the classical rotor model and simulated quantum annealing fail to reproduce the experimental results. We argue that this bolsters the evidence for the reported entanglement.
\end{abstract}
\maketitle
\section{Introduction}
%
The D-Wave processors \cite{Johnson:2010ys,Berkley:2010zr,Dwave} are designed to be physical quantum annealers \cite{Brooke1999} performing adiabatic evolution using programmable superconducting flux qubits. 
These devices have generated a substantial debate in the quantum computing community by laying claim to being the first large scale implementation of a quantum algorithm \cite{EPJ-ST:2015}.  
Much effort has been directed at answering the  fundamental question of whether the D-Wave processors exhibit sufficient ``quantumness" to justify these claims.

Independent verification of the quantumness of the D-Wave devices is challenging in part because of their black-box nature: the user interacts with the device by presenting it with an Ising model problem instance that is programmed as input, and receives a classical  bit string representing the measured state of the qubits in the computational basis as output, at the end of the computation (or quantum annealing run). This state is the device's attempt at finding the ground state of the input Ising problem instance. This input-output interaction mode is clearly not amenable to the usual tests of quantumness emphasizing non-locality \cite{Reichardt:2013db}. Nevertheless, for sufficiently small problems ($\lesssim 20$ qubits), specific instances that emphasize quantum features of the evolution have been designed and found to show strong agreement only with quantum master equations, i.e., open quantum system models, but not with classical models \cite{q-sig,q-sig2,Boixo:2014yu}.  However, for a class of much larger ($> 100$ qubits) random Ising model problems, the device's output exhibited strong correlations \cite{q108,SSSV} with a classical rotor model with Monte Carlo updates due to Shin, Smith, Smolin \& Vazirani (SSSV) \cite{SSSV}, with simulated quantum annealing (SQA) implemented using quantum Monte Carlo \cite{sqa1,Santoro,q108}, and with Parallel Tempering simulations~\cite{martinmayor:15}.
While a detailed study of the excited states and degenerate ground states showed significant deviations between the device and these models \cite{Albash:2014if}, these results nevertheless keep alive the question of whether SQA and the SSSV model provide an effective microscopic description of the device at large numbers of qubits. This question is particularly pertinent since both models are efficiently simulatable on classical computers. Another approach that has been used successfully to model the D-Wave devices is the quantum adiabatic master equation (ME) \cite{ABLZ:12-SI}. The latter is so far the only model that has successfully captured all aspects of the ``quantum signature" experiments reported in Refs.~\cite{q-sig,q-sig2}. Importantly, unlike SQA, this quantum model does not lend itself to an efficient classical simulation. A related master equation (based on the noninteracting blip approximation \cite{Leggett:87}) was successfully used to model collective tunneling in experiments involving a D-Wave device \cite{Boixo:2014yu}.

In contrast to the black-box approach, permitting observations only at the end of each annealing run, recent experiments found evidence of entanglement generated during the evolution of the D-Wave devices from input to output, effectively opening the black box \cite{DWave-entanglement}. Specifically, the experiments reported in Ref.~\cite{DWave-entanglement} 
showed, using qubit tunneling spectroscopy \cite{Berkley:2013bf}, that the measured quantum spectrum and thermal populations are in strong agreement with the quantum spectrum and Gibbs state of the transverse-field Ising Hamiltonian---the Hamiltonian the device is supposed to evolve under.  This, in turn, allowed Ref.~\cite{DWave-entanglement} to demonstrate the existence of entanglement in the device using negativity \cite{Vidal:02a,Love:2007} and an entanglement witness \cite{Terhal:2000qf,Guhne:2009,Spedalieri:2012fk}.

However, an important caveat is that these experiments only provide an indirect way to detect entanglement, since the underlying assumption is that the quantum transverse-field Ising Hamiltonian is an appropriate model of the device. That is, entanglement was detected under the assumption that the measured spectrum and populations arise from transverse-field Ising models whose ground states are already highly entangled, and not from some other (possibly classical) model. This begs the question of whether alternative, classical or semiclassical models would be equally effective at predicting the observed spectrum and thermal state populations. For, if so, the deduced entanglement witness would not be applicable. 

To make this important point clearer, we might formulate it as a ``model loophole," in the tradition of the loopholes associated with nonlocality and Bell inequality tests (see, e.g., Ref.~\cite{Gerhardt:2011nr}). The loophole is simply the fact that entanglement was detected under the assumption that a particular, already-quantum model of the dynamics, is responsible for the observed experimental results.  In an attempt to close this loophole, here we numerically simulate the tunneling spectroscopy experiment using the SSSV model, SQA, and the ME.
We shall demonstrate that SQA and the SSSV model both fail to capture any feature of the entanglement witness experiments reported in Ref.~\cite{DWave-entanglement}, whereas the ME again succeeds. 
In other words, the reported measurements \cite{DWave-entanglement} are consistent with the ME, but inconsistent with SSSV and SQA. If the correct model of the D-Wave device is the ME then the reported measurements imply entanglement. The SQA fails not because it isn't a quantum model, but because it is the incorrect quantum model for the experiments at hand: unlike the ME it does not include a unitary dynamics component, and this dynamics is what is presumably responsible for the experimental observations via an adiabatic connection of eigenstates. Our strategy does of course not rule out the possibility that other classical models, or efficiently simulatable quantum models, might also obtain agreement with the reported measurements. Our results may be seen as an invitation to invent such models, which, if found, would further guide the study of the quantumness question of the D-Wave devices.

The structure of this paper is as follows.  In Section~\ref{sec:review} we review the principle behind the qubit tunneling spectroscopy technique, as well as the theory behind the entanglement measures. In Section~\ref{sec:methods} we describe the simulation methods, in particular the ME, SQA, and the SSSV model. In Section~\ref{sec:experiment} we describe the simulated experiment. We present and discuss our results in Section~\ref{sec:results}, where we demonstrate that only the ME matches the experimental tunneling spectroscopy results.  We conclude in Section~\ref{sec:conclusions}. In the Appendix we demonstrate the robustness of our conclusions to various noise models, and also reject another classical model based on spin dynamics with a friction term.
%
\section{Review} \label{sec:review}
%
\subsection{Qubit tunneling spectroscopy}
%
We briefly review the principle behind qubit tunneling spectroscopy \cite{Berkley:2013bf}, where the goal is to find the energy gaps of the quantum system Hamiltonian $H_{\mathrm{S}}$.  We take $H_{\mathrm{S}}$ to be of the form:
\beq 
\label{eq:S}
H_{\mathrm{S}} = -A \sum_{i =1}^N \sigma_i^x + B H_{\mathrm{IS}} \ ,
\eeq
where $A, B>0$ are constants and 
\beq
H_{\mathrm{IS}} = \sum_{i} h_i \sigma^z_i + \sum_{i<j} J_{ij} \sigma^z_i\sigma^z_j
\eeq
is an Ising Hamiltonian acting only on the system qubits. The $h_i$ and $J_{ij}$ are the local fields and couplings, respectively, and we use $\sigma_i^x$ ($\sigma_i^z$) to denote the Pauli $x$ ($z$) matrix acting on qubit $i$. 

We denote the eigenstates and eigenenergies of $H_{\mathrm{S}}$ by $\{\ket{E_n}\}_{n=1}$ and $\{E_n\}_{n=1}$ respectively with $E_1~\leq~E_2~\leq~\cdots$.  A probe qubit P is coupled to system qubit $1$ to give a system+probe Hamiltonian:
\bes 
\label{eq:S+P}
\begin{align}
H_{\mathrm{S+P}} &= H_{\mathrm{S}}   + B H_{1\mathrm{P}} \ , \\
H_{1\mathrm{P}} &= J_{1\mathrm{P}} \sigma_1^z \sigma_{\mathrm{P}}^z -  J_{1\mathrm{P}} \sigma_1^z -  h_{\mathrm{P}} \sigma_{\mathrm{P}}^z \ .
\end{align}
\ees
An offset local field $\propto -J_{1\mathrm{P}}$ has been applied to qubit $1$ such that, in the eigenenergy subspace where the probe qubit is in the state $\ket{0}$, the eigenstates of the system are given by $\ket{E_n} \otimes \ket{0}$ with energy $E_n - B h_{\mathrm{P}}$ (where the first ket is the state of the system qubits and the second ket is the probe qubit state). When the probe is in the $\ket{1}$ state, the lowest energy state of the system and probe can be written as $\ket{\psi_0} \otimes \ket{1}$ with eigenenergy $\tilde{\epsilon}_0 = \epsilon_0 + B h_{\mathrm{P}}$, where $\ket{\psi_0}$ is the ground state of $H_{\mathrm{S}}-2 B J_{1\mathrm{P}} \sigma_1^z$ with eigenenergy $\epsilon_0$.

Let us assume that the system and probe are initialized in the state $\ket{\psi_0} \otimes \ket{1}$.  Introducing a small transverse field term ($\propto \sigma_\mathrm{P}^x$) for the probe qubit allows for transitions between the states $\ket{\psi_0} \otimes \ket{1}$ and $\ket{E_n} \otimes \ket{0}$.  In an open quantum system we may expect that the dominant process is incoherent tunneling between these two states \cite{Harris:2008lp}.  By tuning the value of $h_{\mathrm{P}}$, we can make the two states degenerate, i.e., $E_n - B h_{\mathrm{P}} = \epsilon_0 + B h_{\mathrm{P}}$, resulting in a resonant peak in the tunneling rate. Since both $B$ and $h_{\mathrm{P}}$ are known, this allows us to solve for differences of the $E_n$, and by finding the locations of the tunneling peaks as a function of $h_\mathrm{P}$, we can map out the quantum spectrum of $H_{\mathrm{S}}$. For example, for a pair of such tunneling peaks at $h_{\mathrm{P}}^{(1)}$ and $h_{\mathrm{P}}^{(2)}$, corresponding to the $n=1$ and $n=2$ energy eigenstates respectively, the energy gap between the eigenstates $\ket{E_2}$ and $\ket{E_1}$ is then given by:
\beq
E_2 - E_1 = 2 B \left(h_{\mathrm{P}}^{(2)} - h_{\mathrm{P}}^{(1)} \right) \ .
\eeq
%
%
\subsection{Equilibrium distribution}
%
Let us assume that to a very good approximation our system only populates the states $\ket{\psi_0} \otimes \ket{1}$ and $\{\ket{E_n} \otimes \ket{0}\}_{n=1}$, such that the populations in these states sum to unity:
\beq 
\label{eq:populations1}
P(\ket{\psi_0} \otimes \ket{1}) + \sum_{n=1} P(\ket{E_n} \otimes \ket{0}) = 1 \ .
\eeq
If we observe that the probe qubit is in the state $\ket{0}$, the system energy eigenstate populations $P(E_n)$ are given by:
\beq
P(E_n) = \frac{ P(\ket{E_n} \otimes \ket{0})}{\sum_{i=1}  P(\ket{E_i} \otimes \ket{0})} = \frac{ P(\ket{E_n} \otimes \ket{0})}{1-P(\ket{\psi_0} \otimes \ket{1}) } \ . 
\eeq
However, if we have tuned $h_\mathrm{P}$ so  that the states $\ket{\psi_0} \otimes \ket{1}$ and $\ket{E_n} \otimes \ket{0}$ are degenerate, and we have waited long enough that their populations have thermalized (i.e., they have equal populations), then $P(\ket{E_n} \otimes \ket{0}) = P(\ket{\psi_0} \otimes \ket{1})$.  Under these assumptions we can find the energy eigenstate populations entirely in terms of the population of the state $\ket{\psi_0} \otimes \ket{1}$:
\beq \label{eq:fraction}
P(E_n)  = \frac{P(\ket{\psi_0} \otimes \ket{1})}{1 - P(\ket{\psi_0} \otimes \ket{1})} \ .
\eeq
We expect these to match the Gibbs state populations, i.e., $P(E_n)  = e^{-\beta H_\mathrm{IS}}/Z$, where $Z$ is the partition function and $\beta = 1/(kT)$ is the inverse temperature.

\subsection{Evidence for entanglement}

In Ref.~\cite{DWave-entanglement}, the authors use the populations found in the ground state $P_1$ and first excited state $P_2$ to construct the density matrix of the system $\rho = \sum_{i=1}^2 P_i \ket{E_i} \bra{E_i}$, which assumes that the off-diagonal components in the energy eigenbasis are zero.  Under this assumption (which, as we show below, agrees with the results of our ME simulations) the authors calculate the negativity \cite{Vidal:02a} for all possible bipartitions $A$ of the system,
\beq \label{eq:negativity}
\mathcal{N}(\rho) = \frac{1}{2} \left( \parallel \rho^{\Gamma_A} \parallel_1 - 1\right) \ ,
\eeq
where $\rho^{\Gamma_A}$ denotes the partial transpose of $\rho$ with respect to the bipartition $A$.  Global entanglement is then defined as the geometric mean of the negativity of all bipartitions \cite{Love:2007}, and was shown to be non-zero in Ref.~\cite{DWave-entanglement}.  

A drawback of this approach is that it assumes the off-diagonal elements of the density matrix vanish. To ensure the robustness of an entanglement conclusion, an entanglement witness was used in Ref.~\cite{DWave-entanglement} (based on the theory formulated in Ref.~\cite{Spedalieri:2012fk}):
\beq
\mathcal{W}_A = \ket{\phi}\bra{\phi}^{\Gamma_A} \ , 
\eeq
where $\ket{\phi}$ is the eigenstate of $\ket{E_0}\bra{E_0}^{\Gamma_A}$ with the most negative eigenvalue. The entanglement witness approach succeeds in certifying entanglement even when the off-diagonal elements of the density matrix are not constrained to vanish, thus extending the range of validity of the presence of entanglement beyond that attainable using the negativity approach. 

Using this entanglement witness, two non-trivial checks were performed.
First, experimental errors in the populations $P_1$ and $P_2$ impose linear constraints on the state $\rho$:
%
\beq
\label{eq:constraints}
P_i - \Delta P_i \leq  \Tr \left[ \rho \ket{E_i}\bra{E_i} \right] \leq P_i + \Delta P_i \ .
\eeq
If $\Tr \left[ \mathcal{W}_A \rho \right] < 0$ for all $\rho$ satisfying the experimental constraints in Eq.~\eqref{eq:constraints}, then entanglement is certified for the bipartition $A$.  Optimizing $\Tr [\mathcal{W}_A \rho]$ subject to the above constraints is an instance of a semidefinite program, a class of convex optimization problems for which efficient algorithms are known. 

Second, uncertainties in the specification of the Hamiltonian in Eq.~\eqref{eq:S}, which can lead to changes in the eigenstates $\ket{E_i}$, were included by adding random perturbations ($10^4$ samples in total) to the Hamiltonian, while ensuring that the maximum of $\Tr \left[ \mathcal{W}_A \rho \right] < 0$ for all perturbations. We independently verify that such noise does not change the presence-of-entanglement conclusion in Appendix~\ref{app:ent-robust}.

\section{Simulation Methods} 
\label{sec:methods}
\begin{figure}[t] 
   \centering
   \includegraphics[width=0.95\columnwidth]{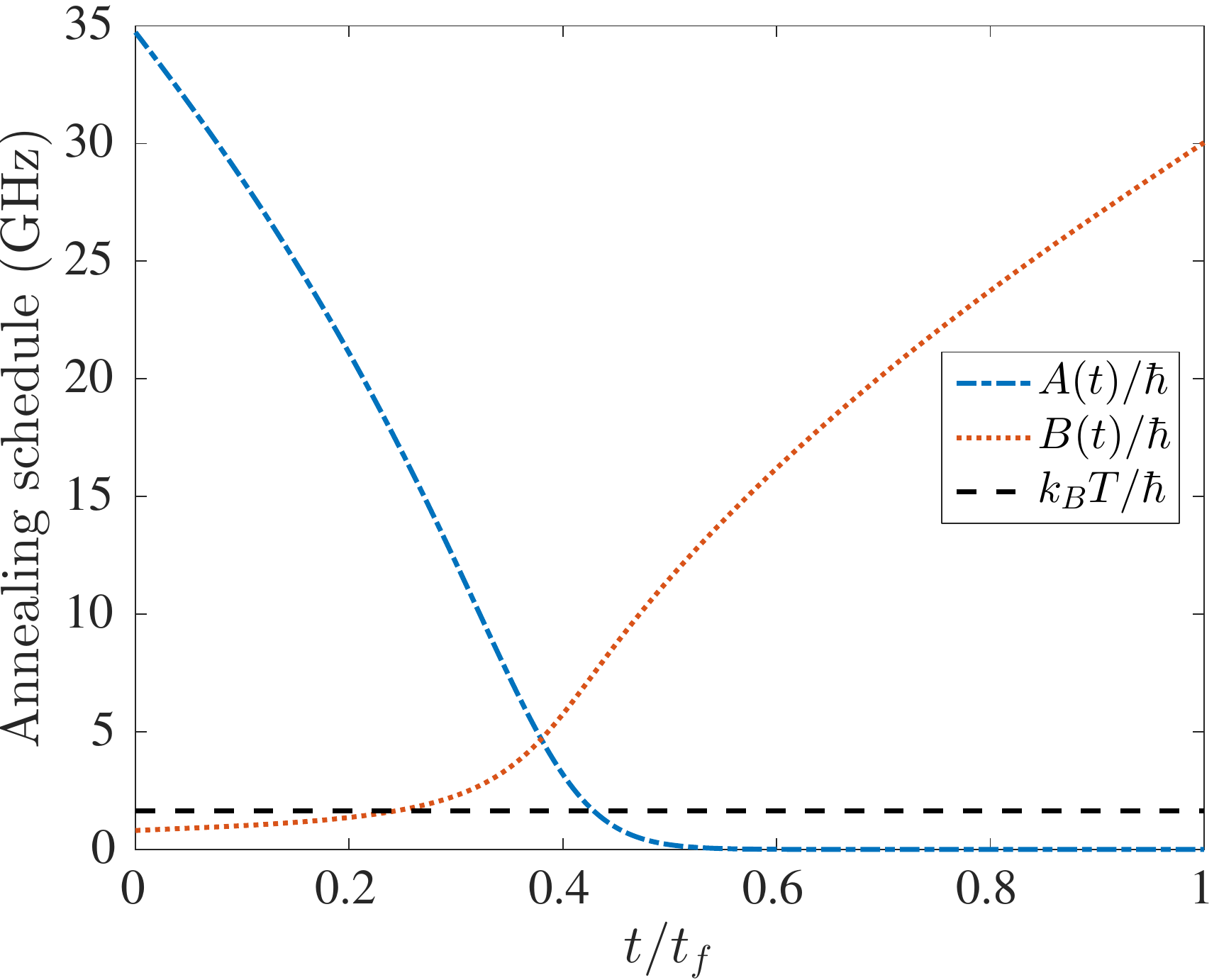} 
   \caption{(Color online) The annealing schedules used in the experiment \cite{DWave-entanglement}. The functional form for $A_\mathrm{S}(s)$ and $A_{\mathrm{P}}(s)$ in Eq.~\eqref{eq:TimeDependentH} is identical to the function $A(s)$ shown.  The dashed line corresponds to the experimental temperature of $12.5$mK.}
   \label{fig:schedule}
\end{figure}

%
The time-dependent Hamiltonian of the experiment is given by
\beq 
\label{eq:TimeDependentH}
H(s) = -A_{\mathrm{S}}(s) \sum_{i =1}^N \sigma_i^x  -A_{\mathrm{P}}(s)\sigma_P^x + B(s) H_{\mathrm{Ising}} \ ,
\eeq
where $s=t/t_f$ is the dimensionless time, and the Ising Hamiltonian $H_{\mathrm{Ising}}$ is given by:
\beq
H_{\mathrm{Ising}} = H_{\mathrm{IS}} + H_{1\mathrm{P}}
\eeq
Just as in Ref.~\cite{DWave-entanglement}, we take $J_{1\mathrm{P}} = -1.8$.  The annealing schedule functions ($A_{\mathrm{S}}(s), A_{\mathrm{P}}(s), B(s)$) are shown in Fig.~\ref{fig:schedule}. These schedules are calculated using rf-SQUID models with independently calibrated qubit parameters \cite{Trevor}, and they correspond to the same schedule used in Ref.~\cite{DWave-entanglement} [see their Fig.~1(d)].  

\subsection{ME} 
\label{sec:ME}
%
We assume that the qubit system is coupled to a bath of independent, thermal, Ohmic oscillators via a dephasing interaction. 
The adiabatic quantum master equation \cite{ABLZ:12-SI} can be used to describe the system evolution in the weak coupling limit, and the evolution of the density matrix is given by:
\begin{eqnarray}
\frac{d}{dt} \rho &=& -\frac{i}{\hbar} \left[H(t), \rho \right] \nonumber \\
&& + \sum_{\alpha=1}^N \frac{g_{\alpha}^2}{\hbar^2}  \sum_{a,b} \Gamma(\omega_{ba}) \left[ L_{ab,\alpha}(t) \rho, \sigma^z_{\alpha} \right] + \mathrm{h.c.} \ ,
\end{eqnarray}
where the index $\alpha$ runs over the qubits, the indices $a,b = 1,\dots,2^N$ run over the instantaneous energy eigenvalues $\eps_a(t)$ of the system Hamiltonian $H(t)$, with Bohr frequencies $\omega_{ba} = [\eps_b(t) - \eps_a(t)]/\hbar$.  The Lindblad operators are given by:
\beq
L_{ab,\alpha}(t) = \bra{ \eps_a(t) } \sigma_\alpha^z \ket{\eps_b(t)} \braket{\eps_a(t)}{\eps_b(t)} \ ,
\eeq
and the function $\Gamma$ encodes the bath correlation function:
\bes \label{eqt:gamma}
\begin{align}
\Gamma(\omega) = & \frac{1}{2} \gamma(\omega) + i S(\omega) \\
= & \frac{\pi \eta e^{-|\omega|/\omega_c}}{1- e^{-\beta \omega}} + i \int_{-\infty}^{\infty} \frac{d \omega'}{2\pi} \gamma(\omega') \mathcal{P} \left(\frac{1}{\omega - \omega'} \right) \ ,
\end{align}
\ees
where $\mathcal{P}$ is the principal value.  
The important free parameters in the ME are the coupling strengths between the system and probe qubits to their respective bosonic baths.  We denote these couplings by $g_{\mathrm{S}}$ and $g_{\mathrm{P}}$ respectively.  
For convenience, we take the system-bath coupling strength $g_{\mathrm{S}}$ to be the same and fixed for all the system qubits, and vary the probe-bath coupling strength $g_{\mathrm{P}}$ relative to $g_{\mathrm{S}}$, expecting $g_{\mathrm{P}}\geq g_{\mathrm{S}}$, since experimentally the probe qubit is operated in a regime where its coupling to the environment is strong (see Supplementary Information of Ref.~\cite{DWave-entanglement}). In the simulations performed in this work, we fix the system-bath coupling strength to:
\beq
g_{\mathrm{S}}^2 \eta / \hbar^2 = 1.2732 \times 10^{-4}  \ ,
\eeq
which is the value used in previous work \cite{q-sig2}.

%
\subsection{SSSV}
The Hamiltonian of the SSSV model \cite{SSSV} is obtained by replacing $\sigma_i^x \mapsto \sin \theta_i$ and $\sigma^z_i \mapsto \cos \theta_i$ in Eq.~\eqref{eq:TimeDependentH}.  The system is evolved by performing Monte Carlo updates on the angles $\theta_i \in [0, \pi]$.  At the end of the evolution, the state is projected onto the computational basis by mapping $\theta_i \leq \pi/2 \to +1$ (state 0) and $\theta_i > \pi/2 \to -1$ (state 1).  This model was derived from first principles using the Keldysh formalism in Ref.~\cite{Crowley:2015fr}.
%
%
\subsection{SQA}
%
We implement a discrete-time version of SQA \cite{sqa1,Santoro,q108}.  At each fixed time $t$ in Eq.~\eqref{eq:TimeDependentH}, Monte Carlo  sampling is performed on the the dual classical spin system with:
\begin{eqnarray} \label{eq:action}
\beta \mathcal{H}_S(t) &=& \frac{\beta}{N_{\tau}} B(t) \sum_\tau \left[ \sum_{i} h_i \mu_{i,\tau} + \sum_{i<j} J_{ij} \mu_{i,\tau} \mu_{j,\tau} \right] \nonumber \\
&&-  \sum_{i,\tau} J_{\perp,i}(t)\mu_{i,\tau} \mu_{i,\tau+1} \ ,
\end{eqnarray}
where $\beta$ is the inverse temperature of the Monte Carlo simulation, $N_\tau$ is the number of Trotter slices used along the Trotter direction, $\mu_{i,\tau}$ denotes the $i$th classical spin on the $\tau$th Trotter slice, and $J_{\tau,i}$ is the nearest-neighbor coupling strength of the $i$th qubit along its Trotter direction and is given by:
\beq
J_{\perp,i}(t) \equiv -  \frac{1}{2}  \ln(\tanh( \beta A_i(t) /N_{\tau})) > 0 \ .
\eeq
In our simulations we fixed $N_{\tau} = 128$ (we checked that increasing it does not alter the results).  

\section{Description of the simulated experiment} 
\label{sec:experiment}
%
The initial Hamiltonian is given by $H(1)$, with $H(s)$ as in Eq.~\eqref{eq:TimeDependentH}.  We choose the initial state of our ME and SQA simulations to be $\ket{\mathbf{1}} \equiv \ket{1}^{\otimes (N+1)}$ (the $N$ system qubits plus the probe qubit); for SSSV we correspondingly choose all the initial angles as $\theta_i = \pi$.  We emphasize that this is not necessarily the ground state of $H(1) = B(1) H_{\mathrm{Ising}}$.  Note that by doing this we are able to skip the state preparation step performed in Ref.~\cite{DWave-entanglement}.  The simulation of the experiment then proceeds as follows:
\begin{enumerate}
\item $B(s)$ and $A_{\mathrm{S}}(s)$ are evolved backward from $s = 1$ to $s = s^{\ast}$ in a time $\tau_1$.  The choice of $s^{\ast}$ determines the quantum Hamiltonian whose spectrum we wish to study.
\item $A_{\mathrm{P}}(s)$ is evolved backward from $s = 1$ to $s = s_{\mathrm{P}}$ in a time $\tau_1$.
\item The system evolves under the constant Hamiltonian $H =  -A_{\mathrm{S}}(s^{\ast}) \sum_{i \in S} \sigma_i^x  -A_{\mathrm{P}}(s_{\mathrm{P}})\sigma_P^x + B(s^{\ast}) H_{\mathrm{Ising}}$ for a ``hold" time $\tau$.  Note that this means that the values of $A$ and $B$ in Eq.~\eqref{eq:S} are given by $A(s^{\ast})$ and $B(s^{\ast})$ respectively.
\item $A_{\mathrm{P}}$ is evolved forward from $s = s_{\mathrm{P}}$ to $s = 1$ in a time $\tau_1$.
\item $B(s)$ and $A_{\mathrm{S}}(s)$ are evolved forward from $s = s^{\ast}$ to $s = 1$ in a time $\tau_1$.
\end{enumerate}
The state of the system qubits and probe qubit are then read. Since the states $\ket{\mathbf{1}}$ and $\ket{\psi_0} \otimes \ket{1}$ are adiabatically connected energy eigenstates, measuring the population change in the $\ket{\mathbf{1}}$ state indicates how much incoherent tunneling \cite{Harris:2008lp} (to the iso-energetic state $\ket{E_n} \otimes \ket{0}$) has occurred at $s^\ast$.  By repeating the experiment for different values of the hold time $\tau$ and recording the probability $P_{\ket{\mathbf{1}}}(\tau,h_{\mathrm{P}},s^{\ast} )$, of observing the $\ket{\mathbf{1}}$ state at the end of the experiment, we can extract the tunneling rate $\Gamma(h_{\mathrm{P}},s^{\ast})$ by fitting  $P_{\ket{\mathbf{1}}}(\tau,h_{\mathrm{P}},s^{\ast} )$ to the function $a +  b e^{-\Gamma(h_{\mathrm{P}},s^{\ast} ) \tau}$.
The experiments are repeated for a range of $h_{\mathrm{P}}$ values in order to find the location of the peaks in $\Gamma$.

\begin{figure}[t] 
  \vspace{0.5cm}
   \hspace{0.2cm}
\includegraphics[width=0.58\columnwidth]{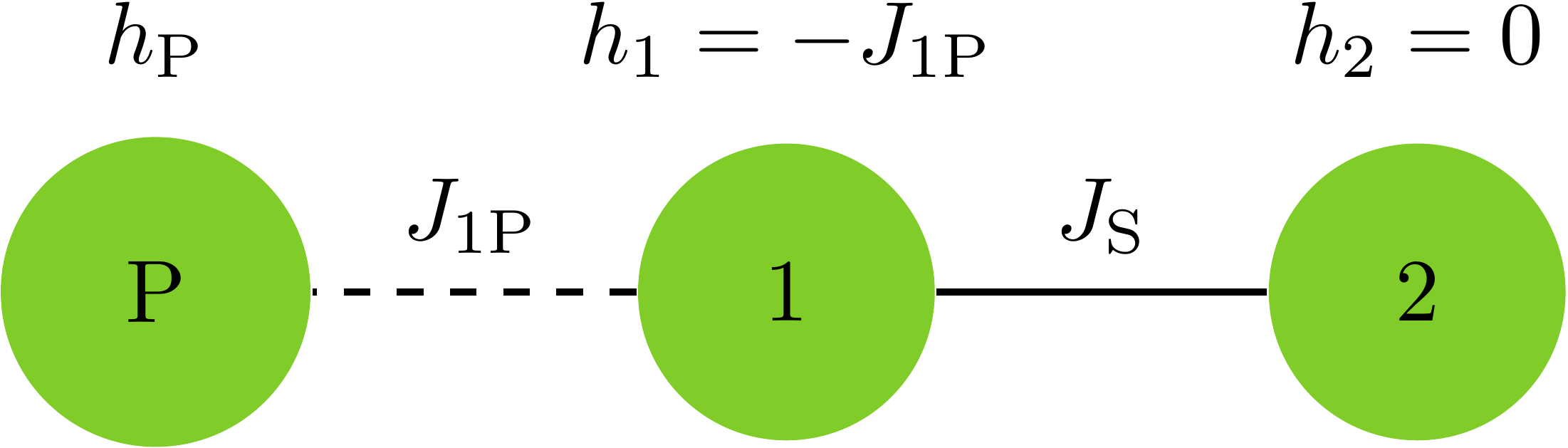} 
   \caption{(Color online) Depiction of the $2+1$-qubit Ising problem.  Qubits are displayed as (labeled) green disks, with their local field value above them.  Couplings are shown as solid lines connecting qubits with their value above the line. Values are picked according to the experiment in Ref.~\cite{DWave-entanglement}, i.e., $J_{\mathrm{S}} = -2.5$, $J_{1 \mathrm{P}} = -1.8$, and $h_{\mathrm{P}}$ is varied.}
   \label{fig:2qubit}
\end{figure}
\begin{figure}[t] 
   \centering
   \includegraphics[width=0.75\columnwidth]{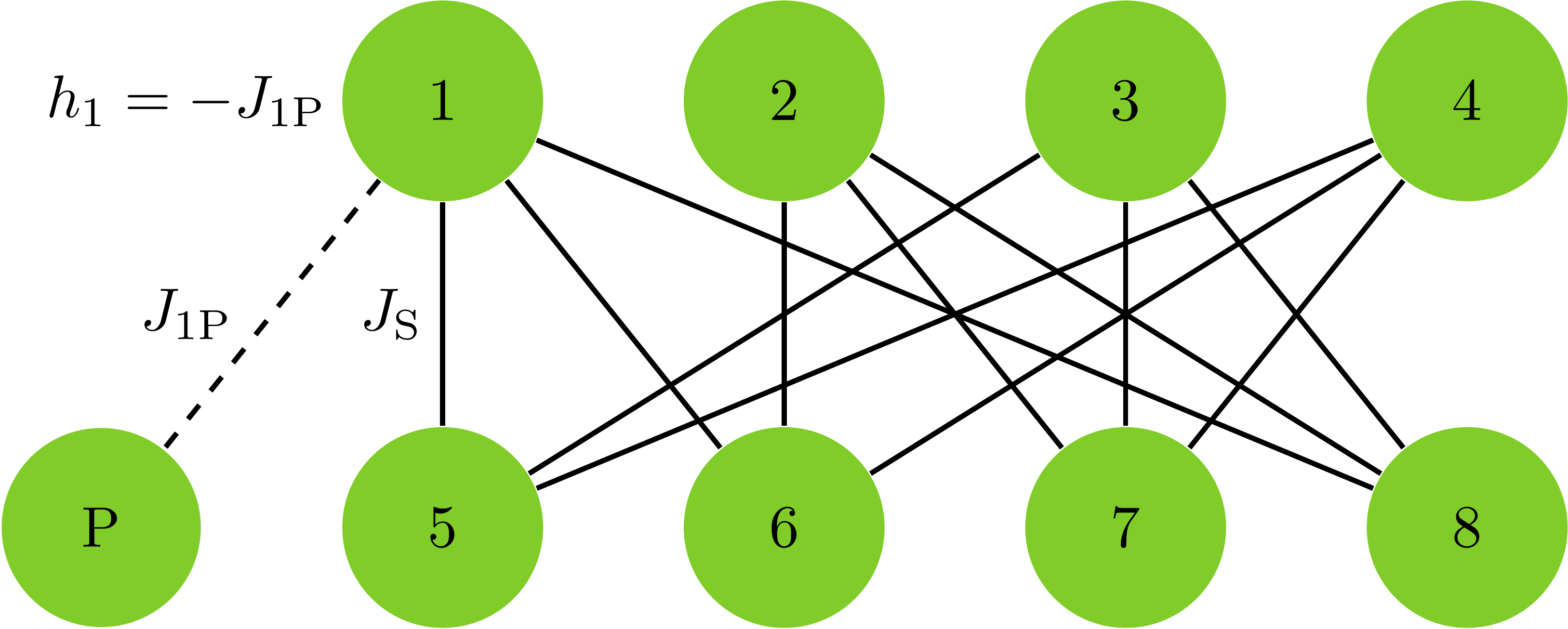} 
   \caption{(Color online) Depiction of the $8+1$-spin Ising problem.  The local fields are zero, and we take the couplings to be $J_\mathrm{S} = -2.5$.  The probe qubit couples as in Fig.~\ref{fig:2qubit}.}
   \label{fig:8qubit}
\end{figure}
\begin{figure}[b] 
   \centering
   \includegraphics[width=0.95\columnwidth]{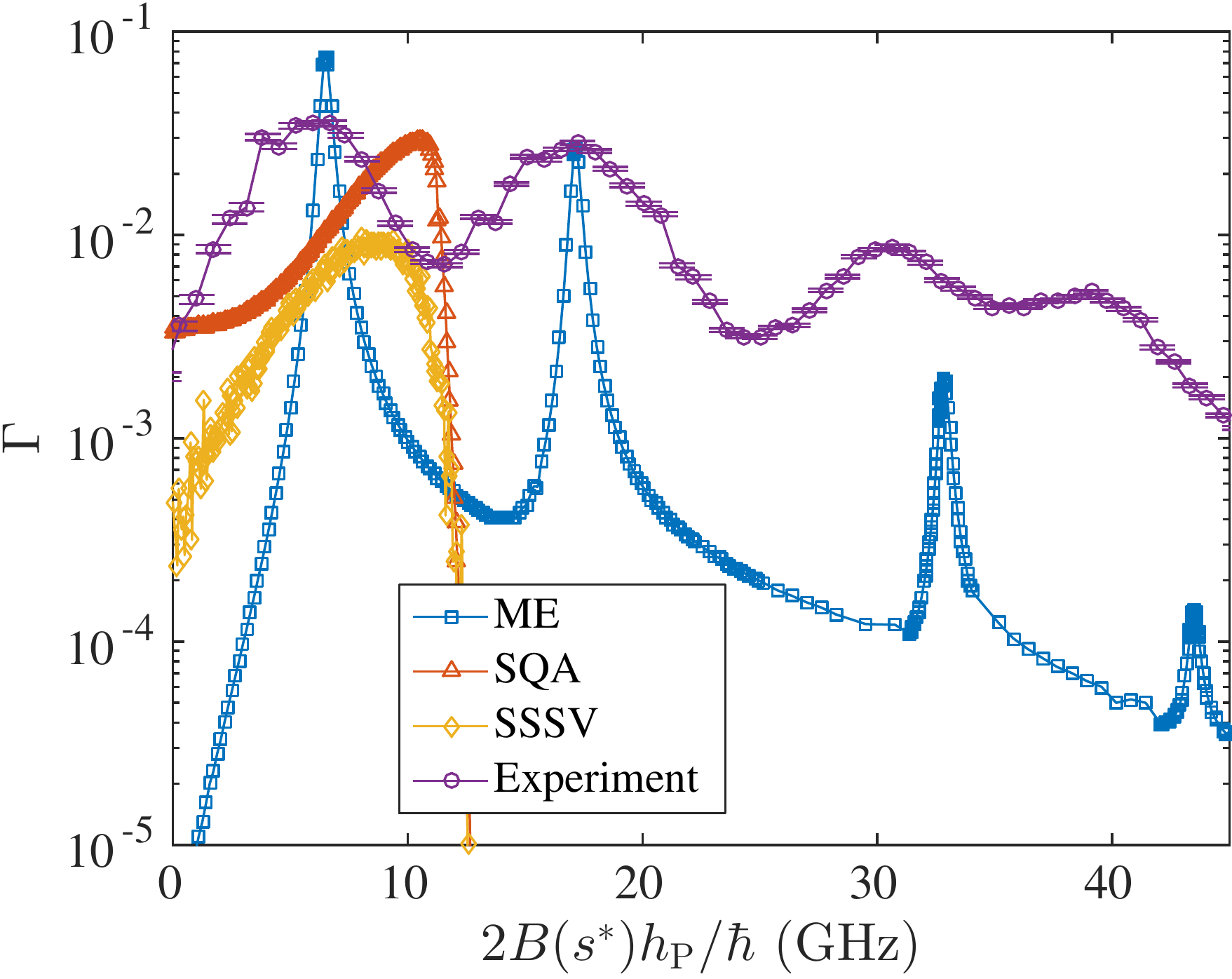}  
   \caption{(Color online) Tunneling rate calculated using the ME, SSSV, and SQA for $s^\ast = 0.339$ and $g_{\mathrm{P}} = 10 g_{\mathrm{S}}$ compared against the experimental results in Ref.~\cite{DWave-entanglement} [see the center figure in their Fig.~8(a)], shifted to the left by $14.5$GHz to align the first peak with the ME results (a constant term in the Hamiltonian present in the experimental setup but not in the ME accounts for this shift).  For the experimental results, 1$\sigma$  error bars are shown.  The two dominant peaks correspond to resonances with the ground state and the first excited state; the magnitude of the peak grows as $g_{\mathrm{P}} \to g_{\mathrm{S}}$ (shown in Appendix \ref{app}), but the peak position remains the same regardless of the value of $g_{\mathrm{P}}$.  For SSSV, we show the results with $T = 12.5$ mK and $\tau_1 = 5$.  $10^6$ runs were performed for each $h_{\mathrm{P}}$ value, and the population of the $\ket{\mathbf{1}}$ state was determined by counting the number of times it occurred in these runs.  For SQA, we show the results with $T = 12.5$ mK and $\tau_1 = 5$.  We performed $10^6$ runs for each $h_{\mathrm{P}}$ value, and the population of the $\ket{\mathbf{1}}$ state was determined by counting the number of times it occurred in these runs.  The tunneling rate for the ME is measured in $\mu s^{-1}$, while for SSSV and SQA it is in inverse sweeps.}
   \label{fig:tunnelingrate}
 \end{figure}
\begin{figure*}[t] 
   \centering
   \subfigure[]{\includegraphics[width=0.95\columnwidth]{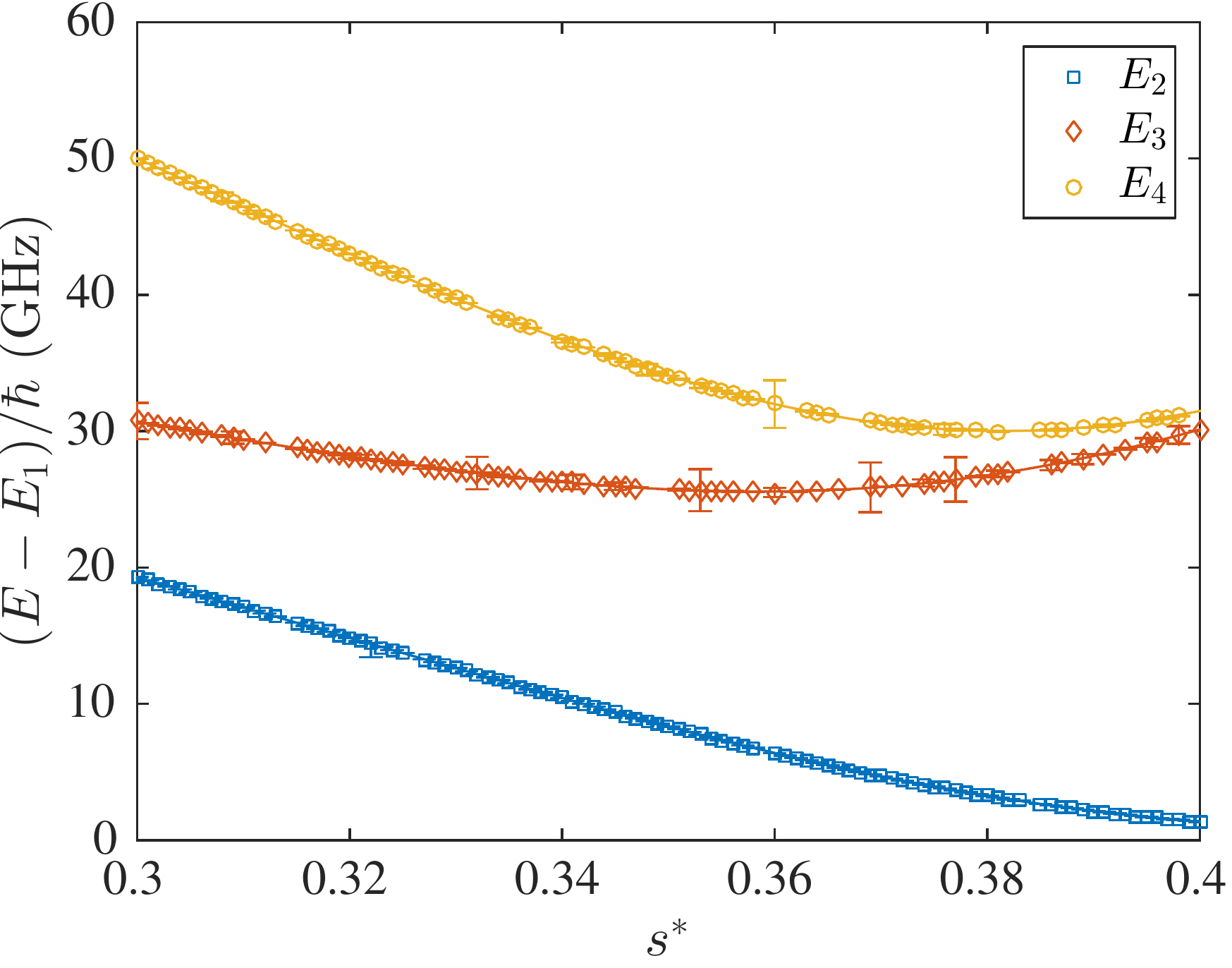}  \label{fig:energylevels} } 
   \subfigure[]{\includegraphics[width=0.95\columnwidth]{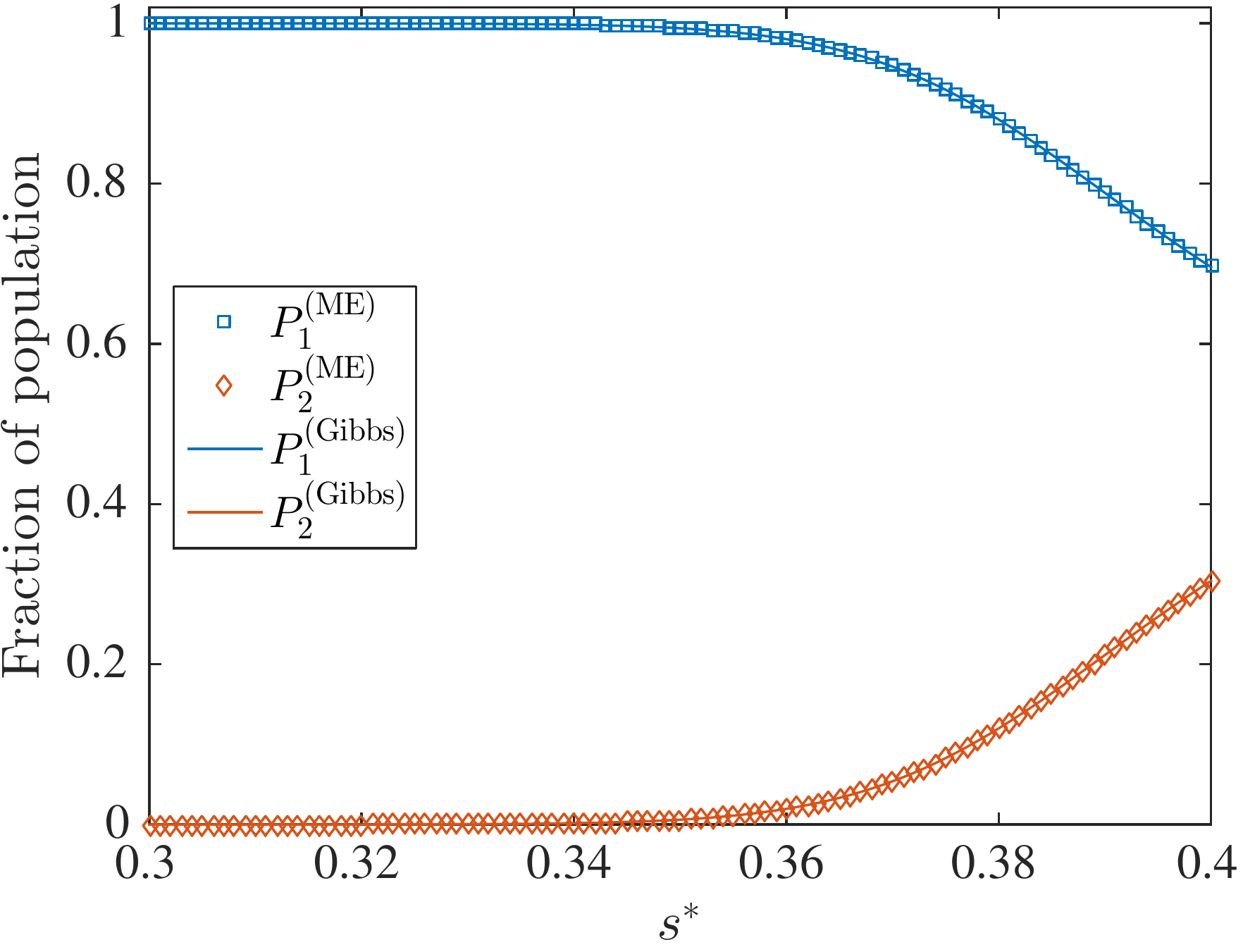}  \label{fig:Gibbs}}
   \caption{(Color online) ME results for the $2+1$ qubit problem.  (a) Energy levels calculated using the tunneling rate peaks via the ME with $g_{\mathrm{P}} = 10 g_{\mathrm{S}}$. $E$ on the vertical axis label is $E_2$, $E_3$, or $E_4$. The solid curves correspond to the theoretical result from diagonalizing $H_{\mathrm{S}}$. Error bars (only visible at a few points) represent $95$\% confidence interval for the fit of the mean for the Gaussian used to estimate the position of the peak in the tunneling rate (see Fig.~\ref{fig:tunnelingrate}). (b) Population fraction as calculated using Eq.~\eqref{eq:fraction} via the ME with $g_{\mathrm{P}} = 10 g_{\mathrm{S}}$, $\tau = 1.5$ms, and a temperature of $12.5$mK.  The solid curves correspond to calculating the Gibbs state associated with $H_{\mathrm{S}}$.  We do not show the population of the higher states since they are $< 10^{-5}$.}
\end{figure*}

The simulations were performed, as in the experiment of Ref.~\cite{DWave-entanglement}, on two system qubits plus one probe qubit, depicted in Fig.~\ref{fig:2qubit}, and eight system qubits plus one probe qubit, depicted in Fig.~\ref{fig:8qubit}.
\section{Results} \label{sec:results}
%

\subsection{ME: $2+1$ qubit results}
%
We first analyze the 2+1 qubit system example studied in Ref.~\cite{DWave-entanglement}, using the adiabatic quantum master equation described in Sec.~\ref{sec:ME}.  We perform the procedure outlined in Sec.~\ref{sec:experiment} with $\tau_1 = 10 \mu s$, and we give an example of the tunneling rate observed at a particular $s^{\ast}$ in Fig.~\ref{fig:tunnelingrate} and compare it to the experimental results of Ref.~\cite{DWave-entanglement}.  We estimate the position of the peak by fitting the data points around the peak with a Gaussian.  We find excellent agreement with the energy spectrum as computed directly from diagonalizing the Hamiltonian, as shown in Fig~\ref{fig:energylevels}.  We then extract the energy eigenstate population distribution using Eq.~\eqref{eq:fraction} at the values of $h_{\mathrm{P}}$ corresponding to the peaks in the tunneling rate, and reproduce the theoretical Gibbs state; see Fig.~\ref{fig:Gibbs}.  Therefore, by reproducing the quantum spectrum and Gibbs populations, the ME is able to reproduce the spectroscopy results of the experiments performed in Ref.~\cite{DWave-entanglement}.

While the relative position of the dominant peaks corresponding to the ground state and first excited state agree very well, we observe that the experimental results are broadened considerably. It is shown in Ref.~\cite{DWave-entanglement} that the experimental broadening is dominated by the linewidth of the probe qubit, which is claimed to be strongly affected by low frequency ($1/f$) noise \cite{Amin}. We attempted to reproduce the broadening using a variety of methods detailed in Appendix~\ref{app}, including the incorporation of low frequency noise, but unfortunately were unable to do so, and this remains an open problem.
Furthermore, the positions of the third and fourth peak corresponding to the third and fourth excited states do not match the experimental result well, but this can be understood as being due to a breakdown of the two-level approximation of the rf-SQUIDs (see the Supplementary Information of Ref.~\cite{DWave-entanglement} and also Ref.~\cite{Amin:13}).
%
\subsection{SSSV: $2+1$ qubit results}
%
We now follow the same procedure using the SSSV model.  Because this model effectively operates in a strong system-bath coupling regime, it thermalizes rapidly, so we are forced to make $\tau_1$ very small (we take each to be only $5$ Monte Carlo sweeps, where a sweep is a complete update of all spins).  Otherwise the SSSV model quickly forgets the initial state before it reaches $s^\ast$, and also forgets the state it relaxed to at $s^{\ast}$ when it returns to $s = 1$.  

We show the tunneling rate calculated using the SSSV model in Fig.~\ref{fig:tunnelingrate}.  Only a single peak is observed over the entire range of $h_{\mathrm{P}}$ values studied, and this peak does not occur at the same position as any of the ME peaks.  It is perhaps not surprising that the SSSV model does not reproduce multiple tunneling peaks, as its energy spectrum is continuous.  The single tunneling peak represents the thermal configuration of the classical rotors at $s^\ast$ that maximally depopulates the $\ket{\mathbf{1}}$ state.  However, as the SQA results will demonstrate next, the failure is ultimately due to the absence of unitary dynamics that adiabatically connects energy eigenstates.

%
\subsection{SQA: $2+1$ qubit results}
%
Unlike the SSSV model's continuous energy spectrum, SQA's spectrum is discrete. However, as we show in Fig.~\ref{fig:tunnelingrate}, SQA also fails to reproduce the experimental tunneling signature on the $2+1$ qubit problem and has a strong similarity to the SSSV model's result (also shown in Fig.~\ref{fig:tunnelingrate}).  We note that SQA and SSSV also correlated strongly on the $108$ random Ising spin problem studied in Ref.~\cite{q108}, as further corroborated and explained in Ref.~\cite{Albash:2014if}.

\begin{figure}[th] 
   \centering
\includegraphics[width=0.75\columnwidth]{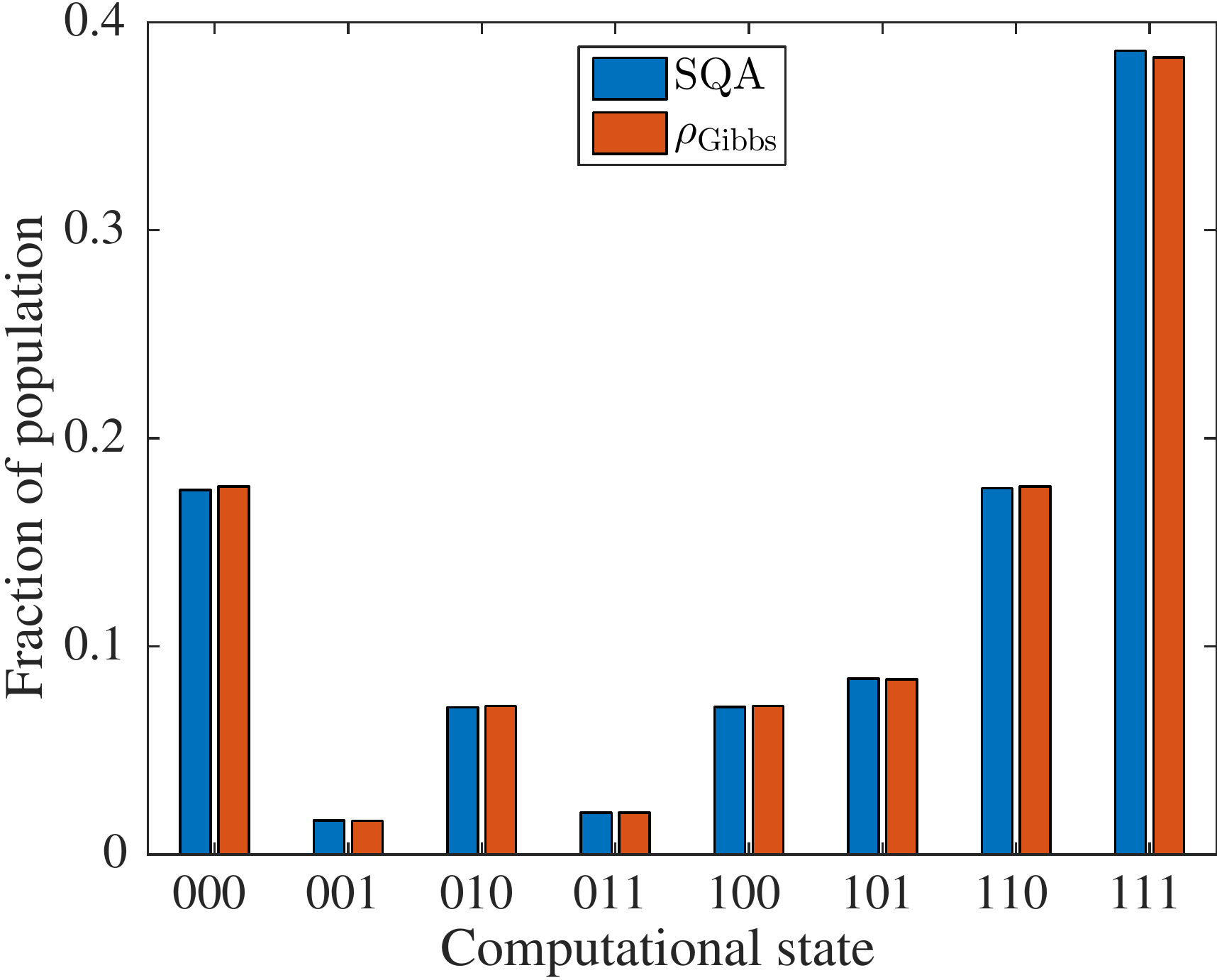}  \label{fig:Pop} 
   \caption{(Color online) The population of the 8 computational states for SQA with $N_\tau = 128$ Trotter slices at 1000 sweeps for $10^6$ repetitions, compared to their population in the quantum Gibbs state.  Both are evaluated at $s^\ast = 0.339$ and $h_P = 1.03$, which is very close to the resonance condition for the states $\ket{E_1} \otimes \ket{0}$ and $\ket{\psi_0} \otimes \ket{1}$.}
   \label{fig:SQAThermal}
\end{figure}

First, we check whether this failure is due to SQA somehow failing to capture the thermal expectation values of observables. However, this is not the case: when we hold SQA at a constant Hamiltonian 
it correctly reproduced the thermal quantum Gibbs state populations for the computational states, as shown in Fig.~\ref{fig:SQAThermal}. Instead, the present failure of SQA is rooted in the absence of unitary dynamics. Namely, there is no adiabatic connection between the initial $\ket{\mathbf{1}}$ state and the state $\ket{\psi_0} \otimes \ket{1}$ at the point $s^{\ast}$.  In order for the arguments presented earlier to work, it is important that during the anneal from $s=1$ to $s^{\ast}$ and $s_{\mathrm{P}}$ and then back to $s=1$, the evolution remain adiabatic with no loss of population from the energy eigenstates, in order for the population of the state $\ket{\mathbf{1}}$ to accurately track the population of the state $\ket{\psi_0} \otimes \ket{1}$. For the ME, this is plausible since all other states with the probe qubit in the $\ket{1}$ state at $s=1$ are at much larger energies, i.e., they correspond to much higher energy eigenstates. Therefore, in a simulation with a unitary dynamics component such as occurs in the ME, we do not expect (nor do we observe) any of these states to be populated such that the only state with the probe qubit in the $\ket{1}$ state is the $\ket{\mathbf{1}}$ state.  However, SQA lacks unitary dynamics, so at best it can provide an adiabatic connection between the thermal state at $s=1$ and at $s= s^{\ast}$, which cannot reproduce the experimental tunneling peak signature.  We also see that SQA does populate other states with the probe qubit in the $\ket{1}$ state besides the $\ket{\mathbf{1}}$ state.  (Note that if we choose to define the tunneling rate in terms of the populations of states with the probe qubit down, our SQA tunneling curves do not change.)

\subsection{ME: 8+1 qubit results}
\begin{figure}[th] 
   \centering
   \includegraphics[width=0.95\columnwidth]{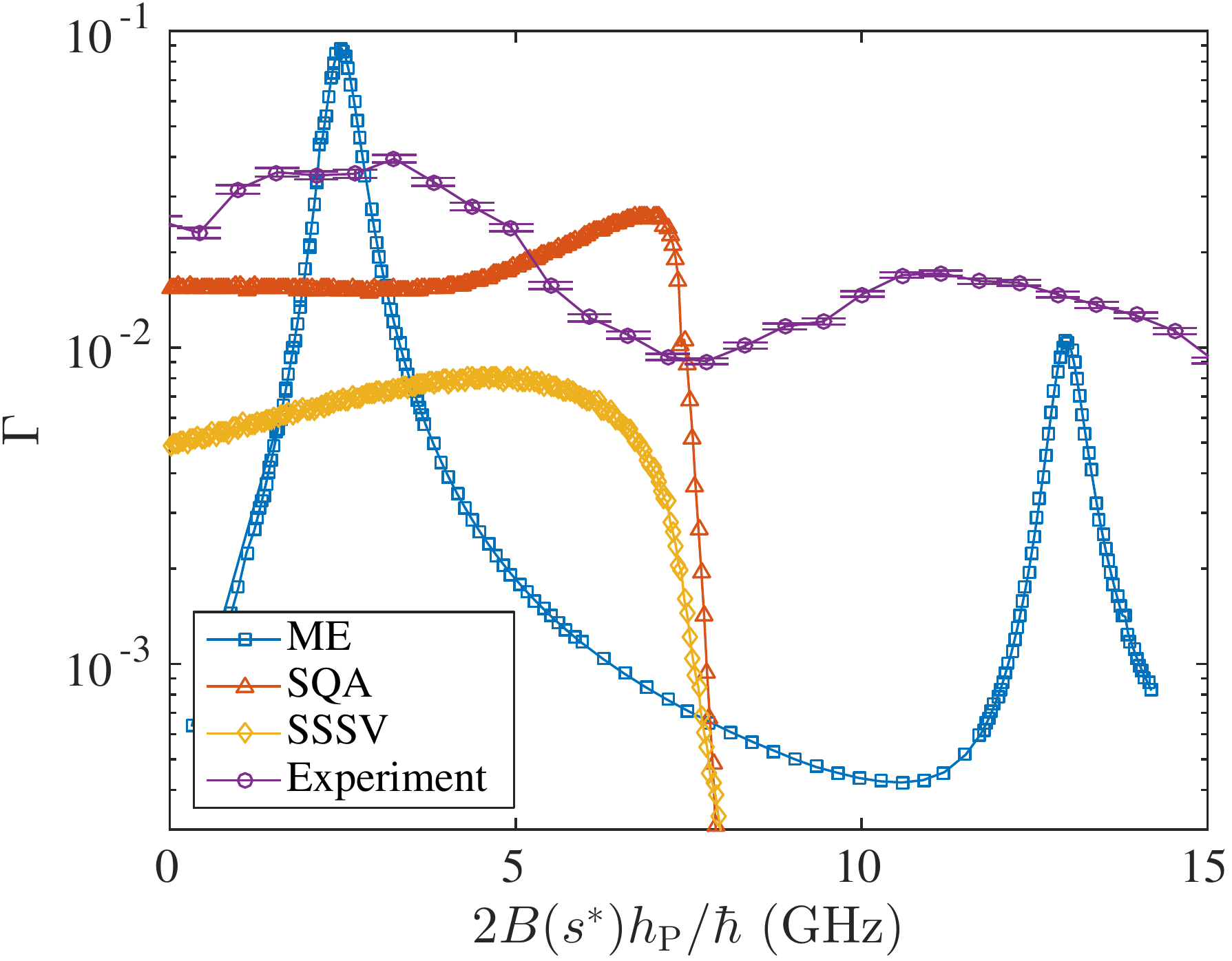} 
   \caption{(Color online) Tunneling rate calculated using the ME, SSSV, and SQA for $s^\ast = 0.284$ as well as the DW experimental results (shifted to align with the first ME peak) on the $8+1$ qubit problem.  For the experimental results, 1$\sigma$  error bars are shown.  We show the results with $T = 12.5$ mK for all simulation methods.  Otherwise, the parameters for all simulation methods are as for the $2+1$ qubit case.  The tunneling rate for the ME is measured in $\mu s^{-1}$, while for SSSV and SQA it is in inverse sweeps.}
   \label{fig:tunnelingrate5}
\end{figure}

%
We extend our analysis to the $8+1$ qubit problem depicted in Fig.~\ref{fig:8qubit}, which was also studied in Ref.~\cite{DWave-entanglement}.  We follow the same method as in the $2+1$ qubit case, the only difference in our numerical simulations being that we truncate the energy spectrum to the lowest $16$ energy eigenstates in order to reduce the computational effort.  
The results are shown in Fig.~\ref{fig:tunnelingrate5}. We continue to find no agreement between the ME and SQA or SSSV. The experimental results exhibit two broad tunneling peaks whose centers match the ME results after a constant shift (as in the $2+1$ qubits case). While the amount of broadening is the same in the $8+1$ and $2+1$ qubit experimental results, the limited range of $h_\mathrm{P}$ values used (extending only up to $15$GHz as opposed to $45$GHz in Fig.~\ref{fig:tunnelingrate}) makes the agreement appear less impressive than in the $2+1$ qubits case.

\begin{figure*}[t] 
   \centering
   \subfigure[]{\includegraphics[width=0.95\columnwidth]{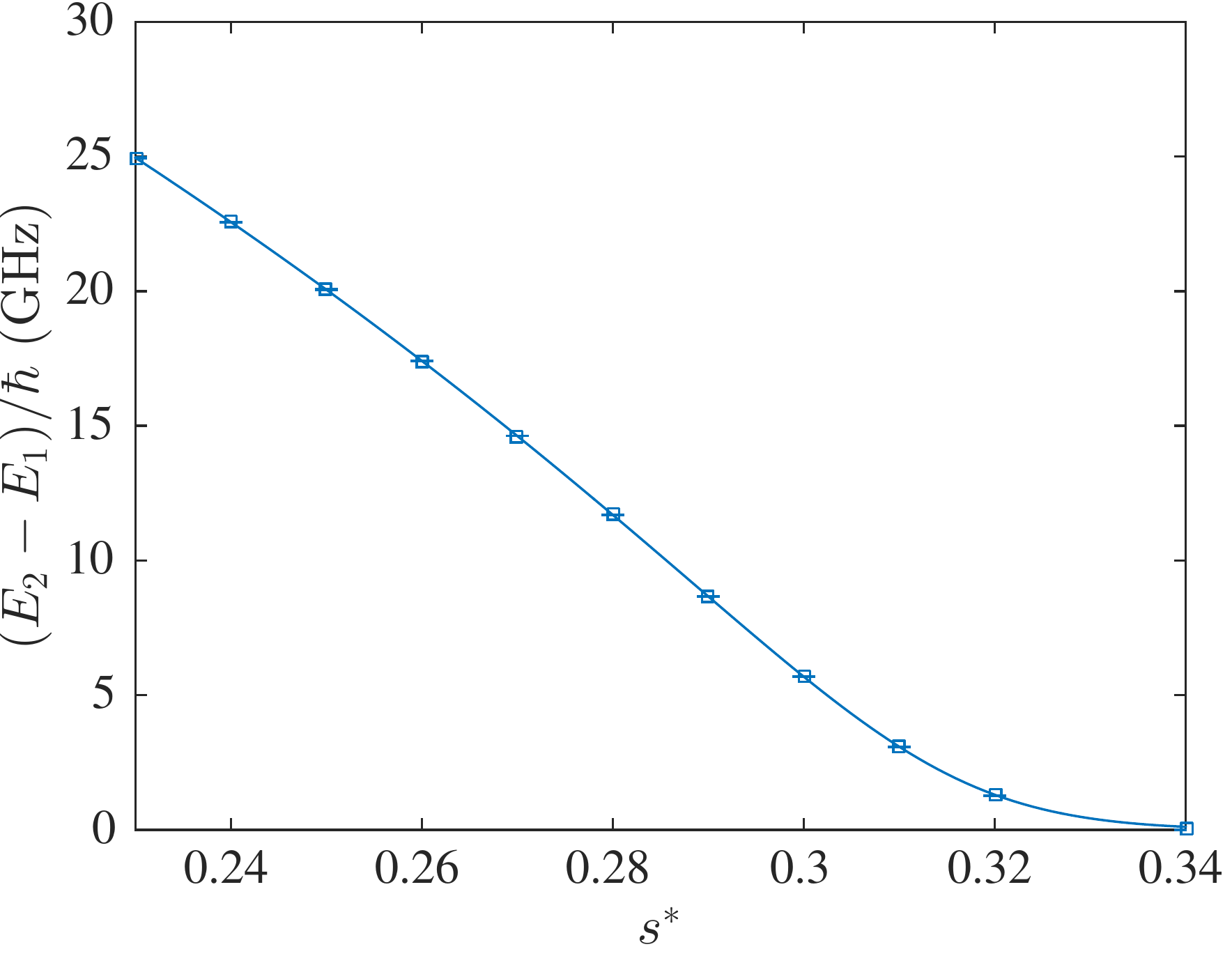}  \label{fig:9Qenergylevels} } 
   \subfigure[]{\includegraphics[width=0.95\columnwidth]{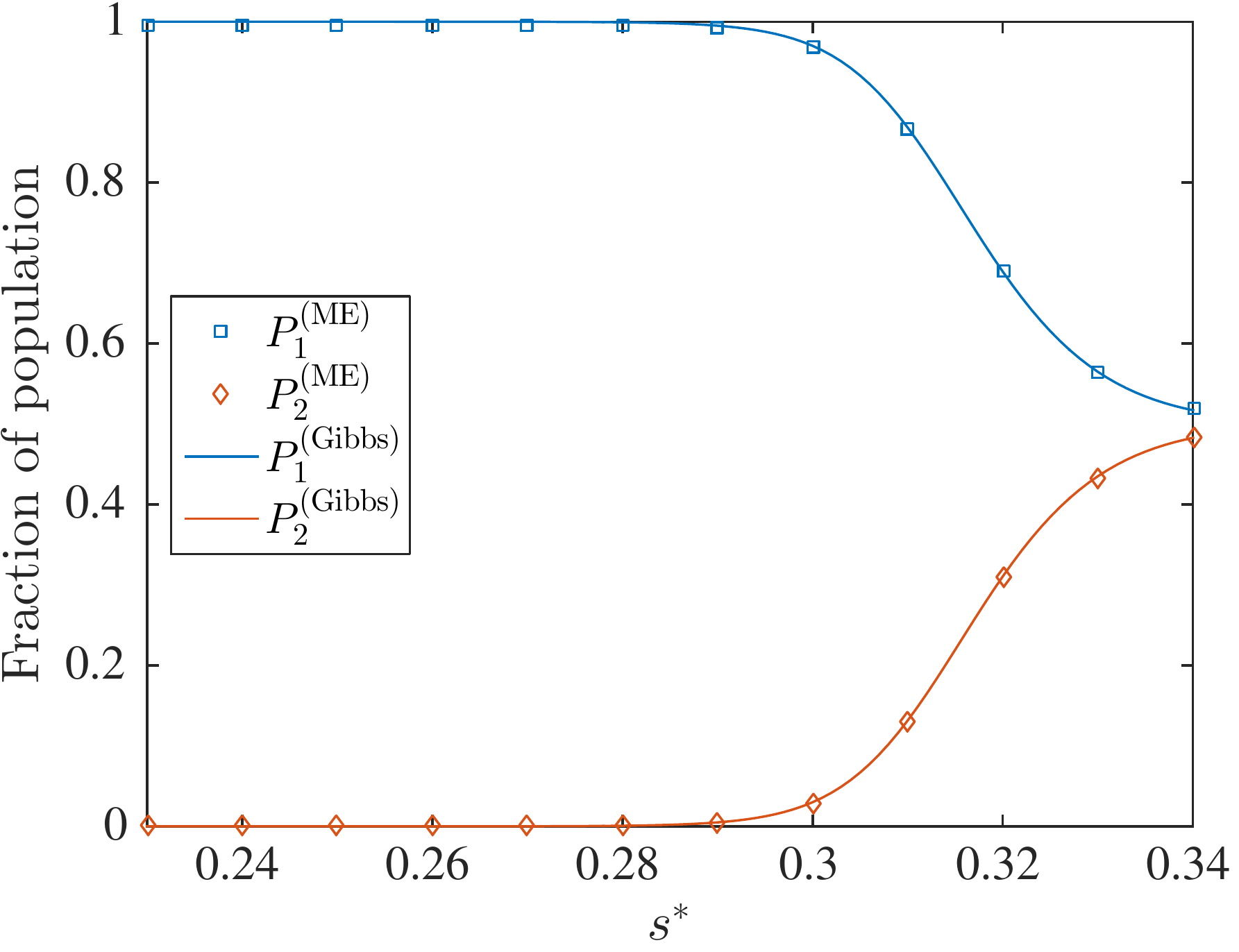}  \label{fig:9QGibbs}}
   \caption{(Color online) Gap and populations ME \textit{vs} Gibbs state results for the $8+1$ qubit problem.  (a) The gap to the first excited state calculated using the tunneling rate peaks via the ME with $g_{\mathrm{P}} = 10 g_{\mathrm{S}}$.  The solid curves correspond to the theoretical result from diagonalizing $H_{\mathrm{S}}$. Error bars (barely visible) represent 95\% confidence interval for the fit of the mean for the Gaussian used to estimate the position of the peak in the tunneling rate (see Fig.~\ref{fig:tunnelingrate}) (b) Population fraction as calculated using Eq.~\eqref{eq:fraction} via the ME with $g_{\mathrm{P}} = 10 g_{\mathrm{S}}$, $\tau = 1.5$ms, and $T = 12.5$mK.  The solid curves correspond to calculating the Gibbs state associated with $H_{\mathrm{S}}$.  We do not show the population of the higher states since they are $< 10^{-5}$. In our simulations, we truncated the energy spectrum to the lowest $16$ energy eigenstates to keep the computational time tractable.  We kept track of the populations along the evolution in order to make sure that this truncation does not lead to a loss of population.  We checked that the results do not change when additional energy levels are included.}
   \label{fig:9QResults}
\end{figure*}

The energy spectrum and the populations as derived from the ME results are shown in Fig.~\ref{fig:9QResults}. The agreement between the ME and the results from diagonalization [in panel (a)] and the Gibbs state [in panel (b)] is again excellent. 

\section{Conclusions} 
\label{sec:conclusions}
%
In this work we set out to check whether the conclusion that entanglement is present during the evolution of $2$- and $8$-qubit experiments using the D-Wave device can be explained using a classical rotor model (SSSV), SQA, or the quantum adiabatic master equation (ME). This question is pertinent since the evidence for entanglement is indirect and assumes a quantum Hamiltonian. We found that neither the SSSV model nor SQA can explain the evidence. The failure of both models can ultimately be attributed to the fact that neither accurately captures the adiabatic connection between energy eigenstates during the annealing evolution. In contrast, the ME reproduces the experimental tunneling spectroscopy signature, suggesting that the underlying transverse field Ising mode Hamiltonian, with its quantized energy spectrum, is an appropriate description, along with the weak coupling approximation used to derive the master equation, at least for the system sizes of up to $8$ qubits we have considered.

Our results support the presence of entanglement in the evolution of the D-Wave device, as concluded in Ref.~\cite{DWave-entanglement}.  Furthermore, the failure of SQA and the SSSV model to capture this result indicates the importance of real-time open system quantum dynamics in supporting this conclusion.  We emphasize that it is important to have both the right thermalization as well as the right dynamics to reproduce the experimental results; for example, we have checked that a Langevin-O(3) model as used in Ref.~\cite{q-sig2} fails to reproduce the experimental results as well (see Appendix~\ref{app:O3}).

Nevertheless, it is important to note that while it becomes computationally prohibitive to use the ME to study systems larger than about $15$ qubits, SSSV and SQA do not have this drawback and seem to capture certain experimental results (in particular the ground state population) at scales of $>100$ qubits \cite{q108,SSSV}.  This tension between the small and large system behaviors remains an open question: does the open quantum system description of the device remain valid at larger sizes, and is the agreement of SQA and the SSSV model with ground state populations due to the fact that they and the open quantum system description yield similar final-time statistics? If so, this could be an artifact of the problem instances studied so far \cite{2014Katzgraber}, or the dominance of thermal noise in the annealing evolution.  Or does the weak-coupling approximation actually break down, so that semi-classical descriptions such as the SSSV model become valid for the dynamics as well?  Further progress in open quantum system modeling and more appropriate benchmarking tests are required in order to address these questions.

\acknowledgements
We thank Trevor Lanting for explaining various details of the experimental procedure, for providing the D-Wave  experimental data, and for providing comments on an early version of the manuscript.  The computing resources were provided by the USC Center for High Performance Computing and Communications.  This research also used resources of the Oak Ridge Leadership Computing Facility at the Oak Ridge National Laboratory, which is supported by the Office of Science of the U.S. Department of Energy under Contract No. DE-AC05-00OR22725.  This research was supported under ARO grant number W911NF-12-1-0523, under ARO MURI Grant No. W911NF-11-1-0268, and under NSF grant number DMS-1529079.

%

\newpage

\appendix


\section{Robustness of the entanglement}
\label{app:ent-robust}
\begin{figure}[h] 
   \centering
   \includegraphics[width=0.85\columnwidth]{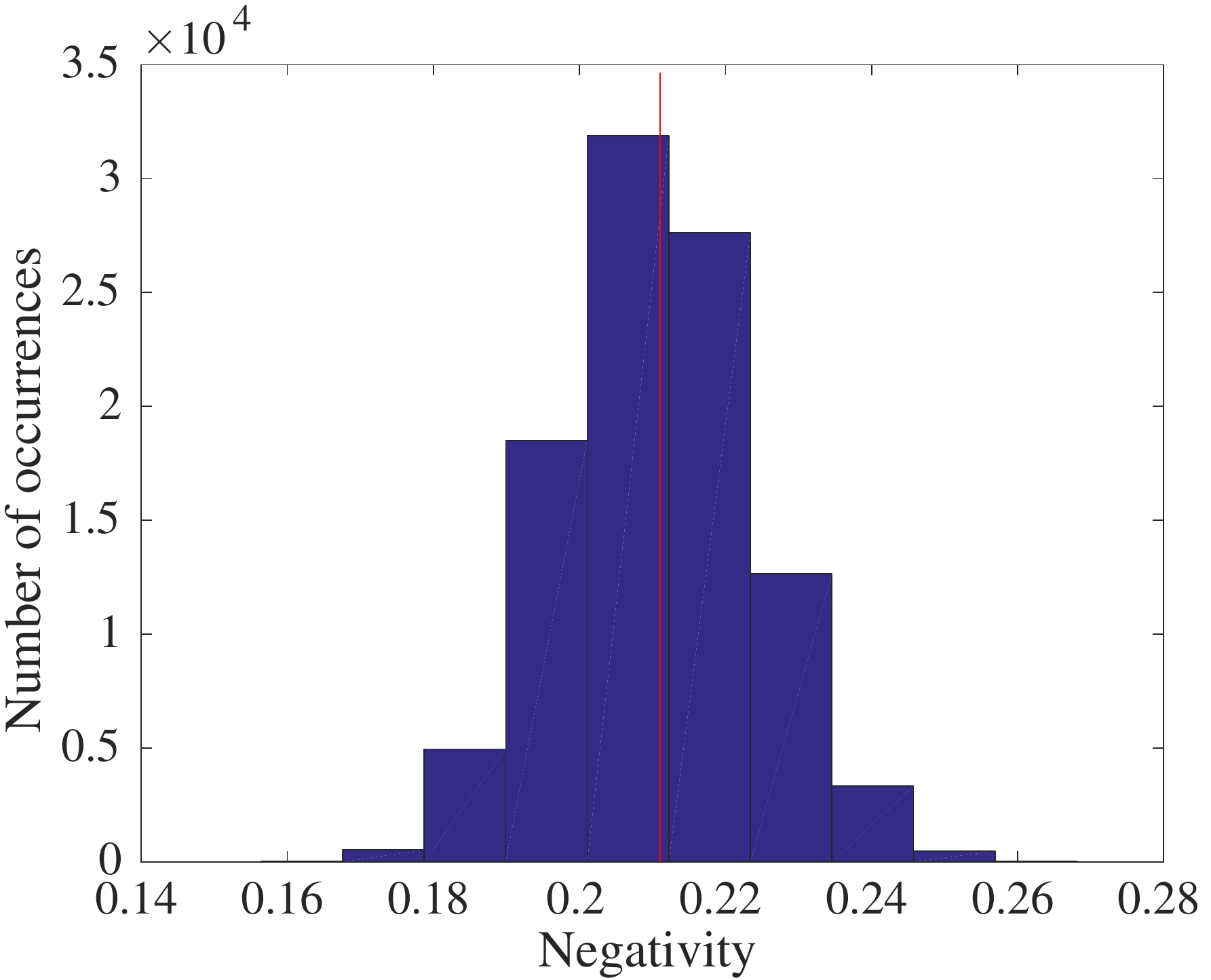}  
   \caption{(Color online) Histogram of the negativity of the Gibbs state associated with the $2$-qubit Hamiltonian of Eq.~\eqref{eq:S} when a noise term of the form given in Eq.~\eqref{eqt:HNoise} is included.  The red vertical line at $0.211$ is the negativity for the noiseless case.  A total of $10^5$ noise instances are shown.}
   \label{fig:HistNegativity}
 \end{figure}

We consider adding Gaussian noise to all the fields and couplings in the $2$-qubit system Hamiltonian [Eq.~\eqref{eq:S}], i.e., we include an additional Hamiltonian of the form:
\begin{eqnarray} \label{eqt:HNoise}
H_{\mathrm{Noise}}(s) &=& A(s) \left( \alpha_{1} \sigma_1^x + \alpha_{2} \sigma_2^x \right) \nonumber \\
&& + B(s) \left( \alpha_{3} \sigma_1^Z + \alpha_{4} \sigma_2^Z + \alpha_{5} \sigma_1^Z \sigma_2^Z  \right) ,
\end{eqnarray}
where $\alpha_i \sim \mathcal{N} (0, 0.1)$.  This choice is well above the typical $\sigma = 0.05$ ICE associated with the D-Wave Two processors.  We show in Fig.~\ref{fig:HistNegativity} that for the $10^5$ noise samples generated, the Gibbs state never has zero negativity [as defined in Eq.~\eqref{eq:negativity}].  Therefore, such a noise model is unlikely to invalidate the presence-of-entanglement conclusion of Ref.~\cite{DWave-entanglement}.
 
\section{Extended noise models}
\label{app}
\begin{figure}[ht] 
   \centering
   \includegraphics[width=0.85\columnwidth]{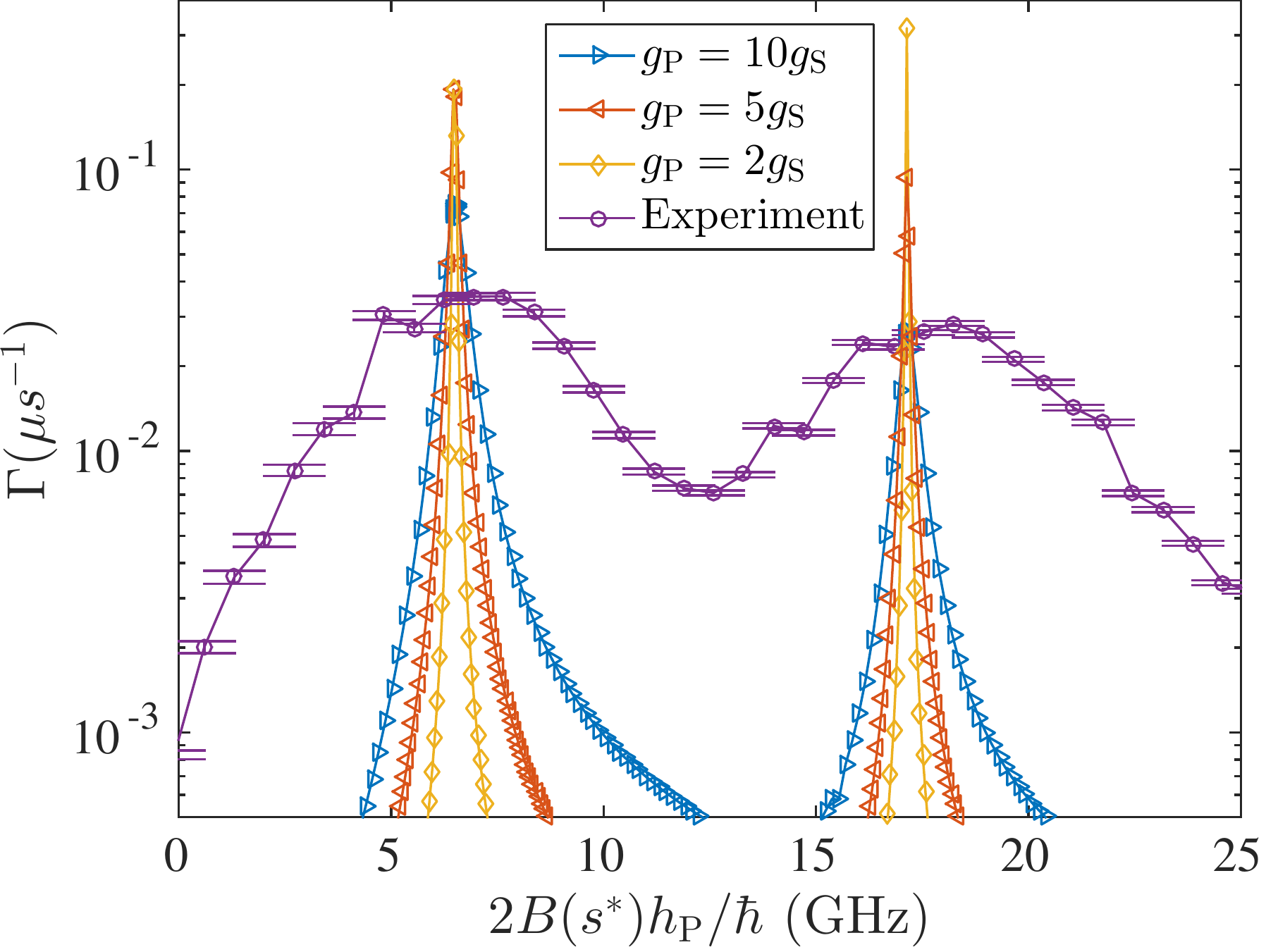}   
     \caption{(Color online) Tunneling rate for the $2+1$ qubit problem calculated using the ME for $s^\ast = 0.339$ and variable $g_{\mathrm{P}}/g_{\mathrm{S}}$ compared against the experimental results in Ref.~\cite{DWave-entanglement}.}
   \label{fig:tunnelingrategS}
 \end{figure}

While the ME is able to reproduce the positions of the tunneling rate peaks (see Figs.~\ref{fig:tunnelingrate} and \ref{fig:tunnelingrate5}), the experimental results show a broadening of the peaks that is absent in the ME results, indicating that an important noise source is missing from the ME simulations. In the Supplementary Information of Ref.~\cite{DWave-entanglement}, it is made clear that the line shape of the first peak (corresponding to the ground state) in the two-qubit and eight-qubit systems are almost identical to the probe qubit's macroscopic resonant tunneling (MRT) profile.  This suggests that the broadening observed is from noise on the probe qubit and not the multi-qubit system.  In our simulations, we have attempted to model this by increasing the relative system-bath coupling strength between the probe and system qubits.  We show in Fig.~\ref{fig:tunnelingrategS} that the broadening induced by this method is not sufficient to explain the observed broadening.

 \begin{figure}[ht] 
   \centering
   \includegraphics[width=0.85\columnwidth]{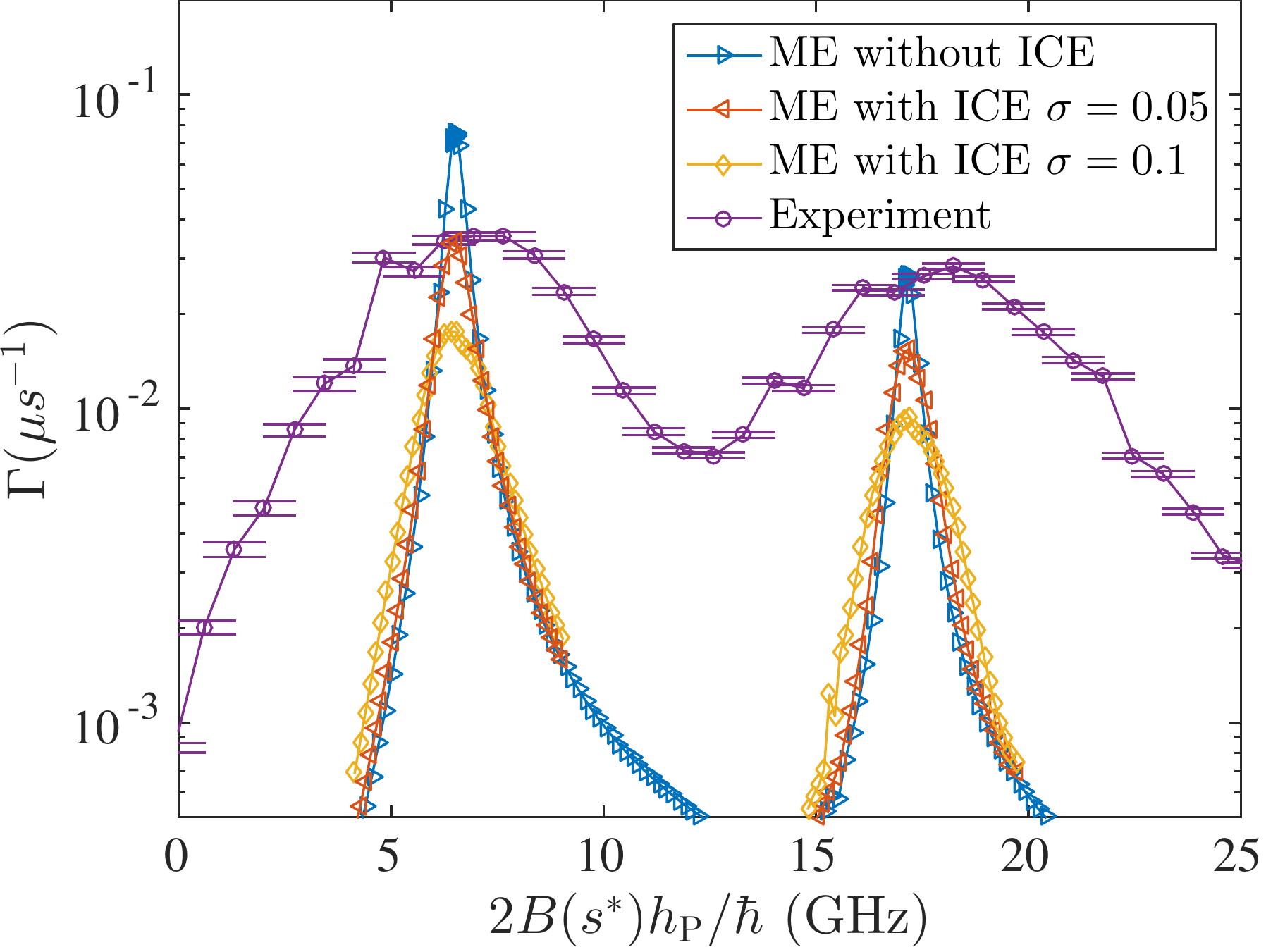}  
   \caption{(Color online) Tunneling rate for the $2+1$ qubit problem calculated using the ME for $s^\ast = 0.339$ and $g_{\mathrm{P}} = 10 g_{\mathrm{S}}$ with $\sigma = 0.05$ and $\sigma = 0.1$ ICE on the probe qubit only and without ICE compared against the experimental results in Ref.~\cite{DWave-entanglement}.}
   \label{fig:tunnelingrateProbeICE}
 \end{figure}
 \begin{figure}[ht] 
   \centering
   \includegraphics[width=0.85\columnwidth]{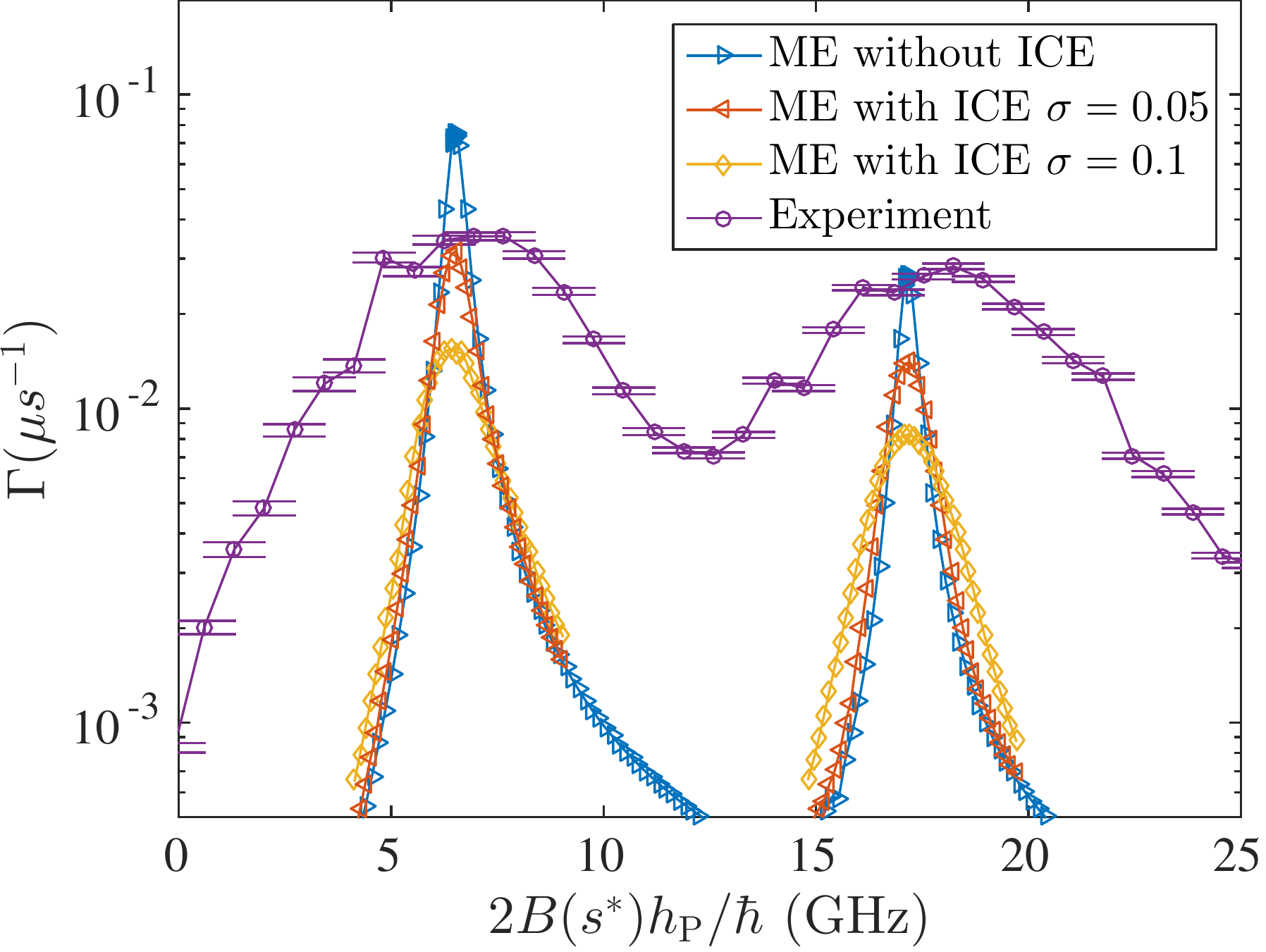}  
   \caption{(Color online) Tunneling rate for the $2+1$ qubit problem calculated using the ME for $s^\ast = 0.339$ and $g_{\mathrm{P}} = 10 g_{\mathrm{S}}$ with $\sigma = 0.05$ and $\sigma = 0.1$ ICE on all qubits and without ICE compared against the experimental results in Ref.~\cite{DWave-entanglement}.}
   \label{fig:tunnelingrateICE}
 \end{figure}
  \begin{figure}[t] 
   \centering
   \includegraphics[width=0.85\columnwidth]{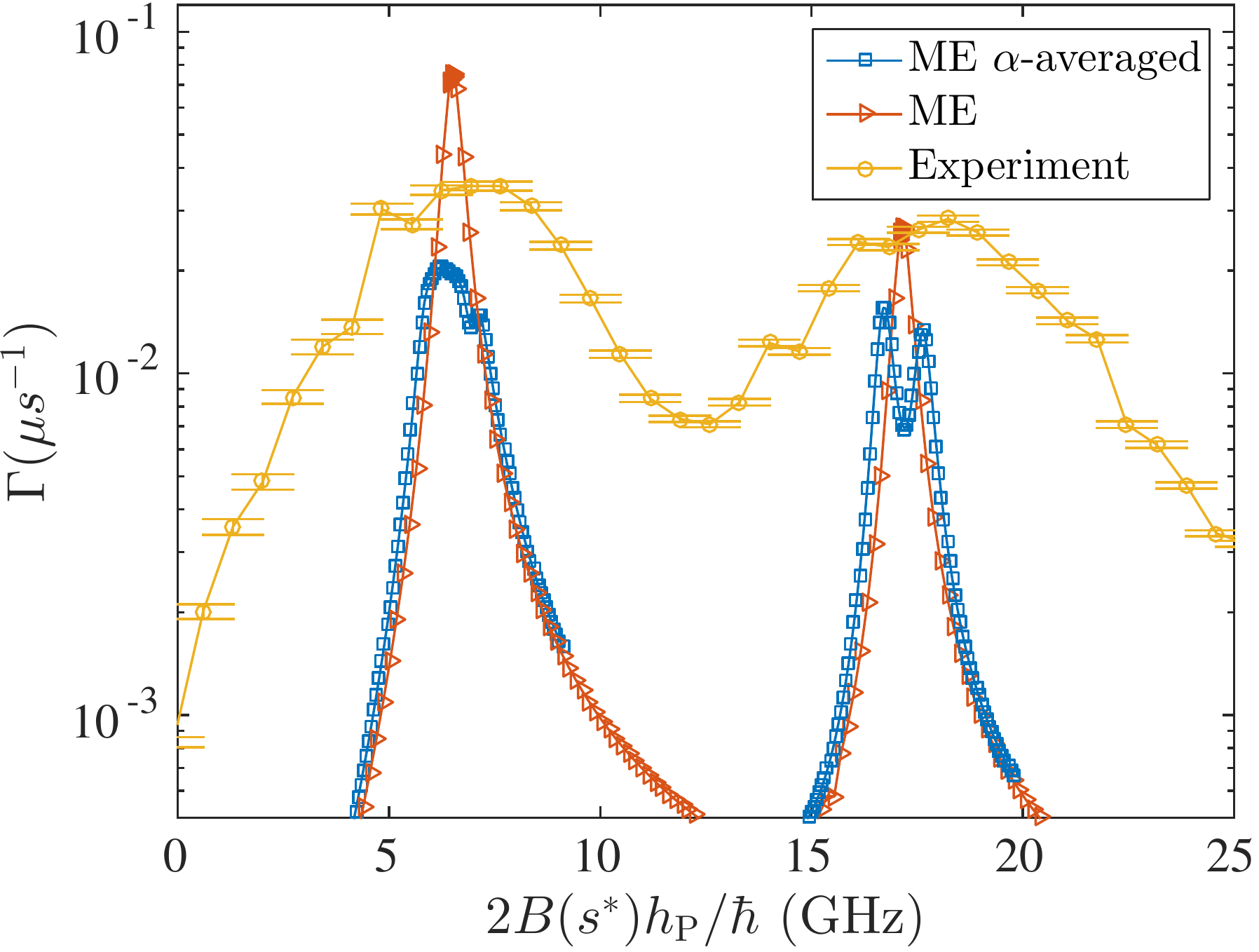}  
   \caption{(Color online) Tunneling rate for the $2+1$-qubit problem calculated using the ME for $s^\ast = 0.339$ and $g_{\mathrm{P}} = 10 g_{\mathrm{S}}$ averaged over $\alpha$ [see Eq.~\eqref{eqt:alpha}] compared against the experimental results in Ref.~\cite{DWave-entanglement}.}
   \label{fig:tunnelingrateAlpha}
 \end{figure}
We attempt therefore to reproduce this broadening using several other possible noise models. Our noise model choices are phenomenological, and are designed to test whether such modifications are sufficient to broaden the tunneling peaks.  First, we consider introducing Gaussian noise 
on the Ising local fields and couplings:
     \begin{figure}[t] 
   \centering
   \includegraphics[width=0.85\columnwidth]{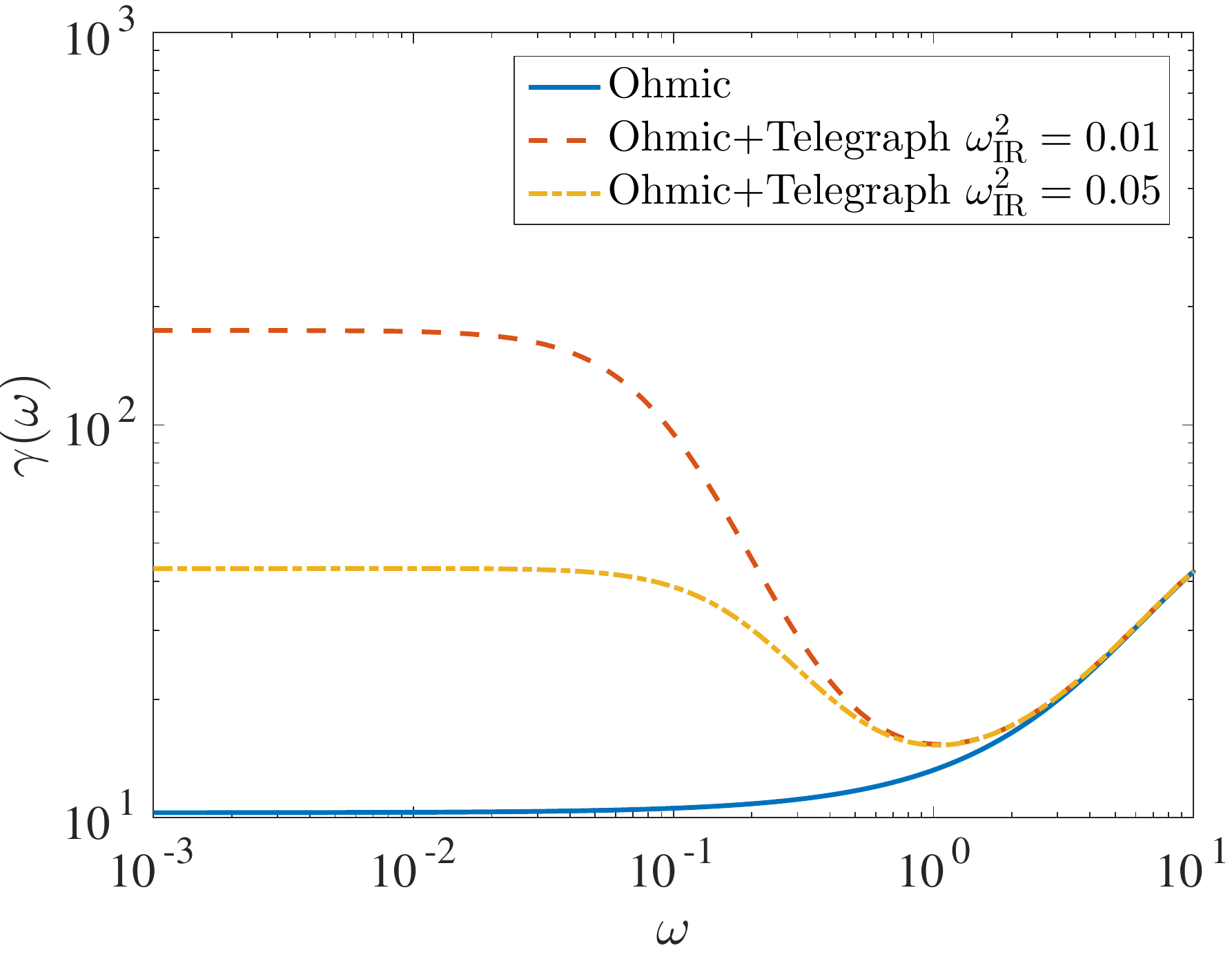}     \caption{(Color online) The modifications due to the addition of telegraph noise [Eq.~\eqref{eqt:gamma2}] to the purely Ohmic spectral density function $\gamma(\omega)$.}
   \label{fig:gammafunction}
 \end{figure}
\beq
J_{ij} \to J_{ij} + \mathcal{N}(0, \sigma) \ , \quad h_i \to h_i + \mathcal{N}(0, \sigma)
\eeq
This is a relevant source of noise for the D-Wave processors, often referred to as internal control error (ICE) \cite{King:2014uq}.  We run the ME with $1000$ noise realizations for each applied (ideal) bias $h_p$.  We then average the observed population in the $\ket{\mathbf{1}}$ state over these 1000 realizations and then extract the tunneling rate associated with the applied (ideal) bias $h_p$.  We first consider the case where the noise is purely on the local field and coupling of the probe qubit.  We show in Fig.~\ref{fig:tunnelingrateProbeICE} results with $\sigma = 0.05$ and $\sigma=0.1$, well above the typical $\sigma = 0.05$ ICE associated with the D-Wave Two ``Vesuvius" processors.  While the introduction of ICE in the simulations reduces the height of the tunneling-rate peaks and broadens them slightly, it is insufficient to account for the broadening observed in the experiment.  Similar results are achieved if ICE is introduced on all qubits (see Fig.~\ref{fig:tunnelingrateICE}), suggesting this noise model is not sufficient to explain the broadening obesrved.

    \begin{figure}[t] 
   \centering
   \includegraphics[width=0.85\columnwidth]{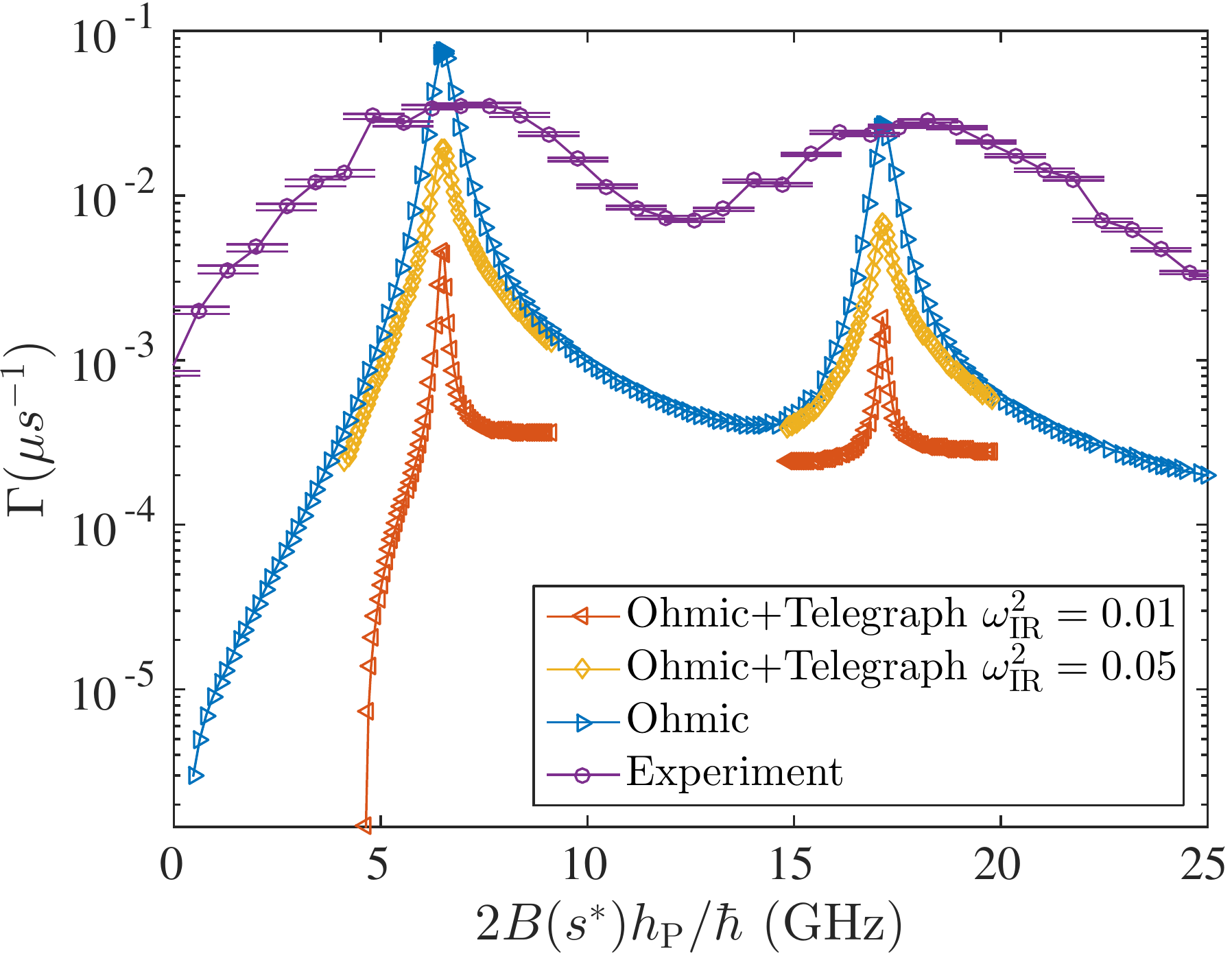}  
   \caption{(Color online) Tunneling rate for the $3$ qubit problem calculated using the ME for $s^\ast = 0.339$ and $g_{\mathrm{P}} = 10 g_{\mathrm{S}}$ using purely Ohmic and Ohmic plus telegraph spectral noise compared against the experimental results in Ref.~\cite{DWave-entanglement}.}
   \label{fig:Ohmic}
 \end{figure}
 
 A more restrictive (and less physically reasonable) noise model is:
 \beq \label{eqt:alpha}
 h_{i} \to h_{i}  + \alpha  \ , \quad i = 1,2 \ , 
 \eeq
where we average over three $\alpha \in \{-0.1,0,0.1\}$ values.  The result is shown in Fig.~\ref{fig:tunnelingrateAlpha}. While this does not sufficiently broaden the peaks, it does cause the appearance of peak splitting, which is a feature of the experimental results.

Finally, we modify the spectral density $\gamma(\omega)$  in Eq.~\eqref{eqt:gamma} to include a low frequency component. We consider a form of telegraph noise \cite{PhysRevApplied.3.044009}, which we model as
 \beq \label{eqt:gamma2}
 \gamma_{\mathrm{tel}}(\omega) = \frac{\omega}{1 - e^{-\beta \omega}} \frac{1}{\omega^2 + \omega_{\mathrm{IR}}^2} \ ,
 \eeq
in order to satisfy the KMS condition \cite{KMS}. We set $\omega_{\mathrm{IR}}^2=0.01, 0.05$ GHz$^2$(this choice is motivated by numerical stability; making $\omega_{\mathrm{IR}}$ and $\omega_{\mathrm{IR}}$ smaller results in a significant slowdown of our simulations), and assume that this contribution to $\gamma(\omega)$ has the same coupling strength $g$ as the Ohmic component.  We show in Fig.~\ref{fig:gammafunction} how this modifies the purely Ohmic case.  As we show in Fig.~\ref{fig:Ohmic}, these modifications are counterproductive and act to narrow the peak. We also included $1/f$ noise [with a spectral density of the form $\gamma_{1/f}(\omega) =  \frac{2 \pi e^{-|\omega|/\omega_c}}{1 - e^{-\beta \omega}} \frac{ |\omega|}{\omega^2 + \omega_{\mathrm{IR}}^2}$], but this did not reproduce the experimental broadening either (not shown).

     \begin{figure*}[ht] 
   \centering
   \subfigure[]{\includegraphics[width=0.85\columnwidth]{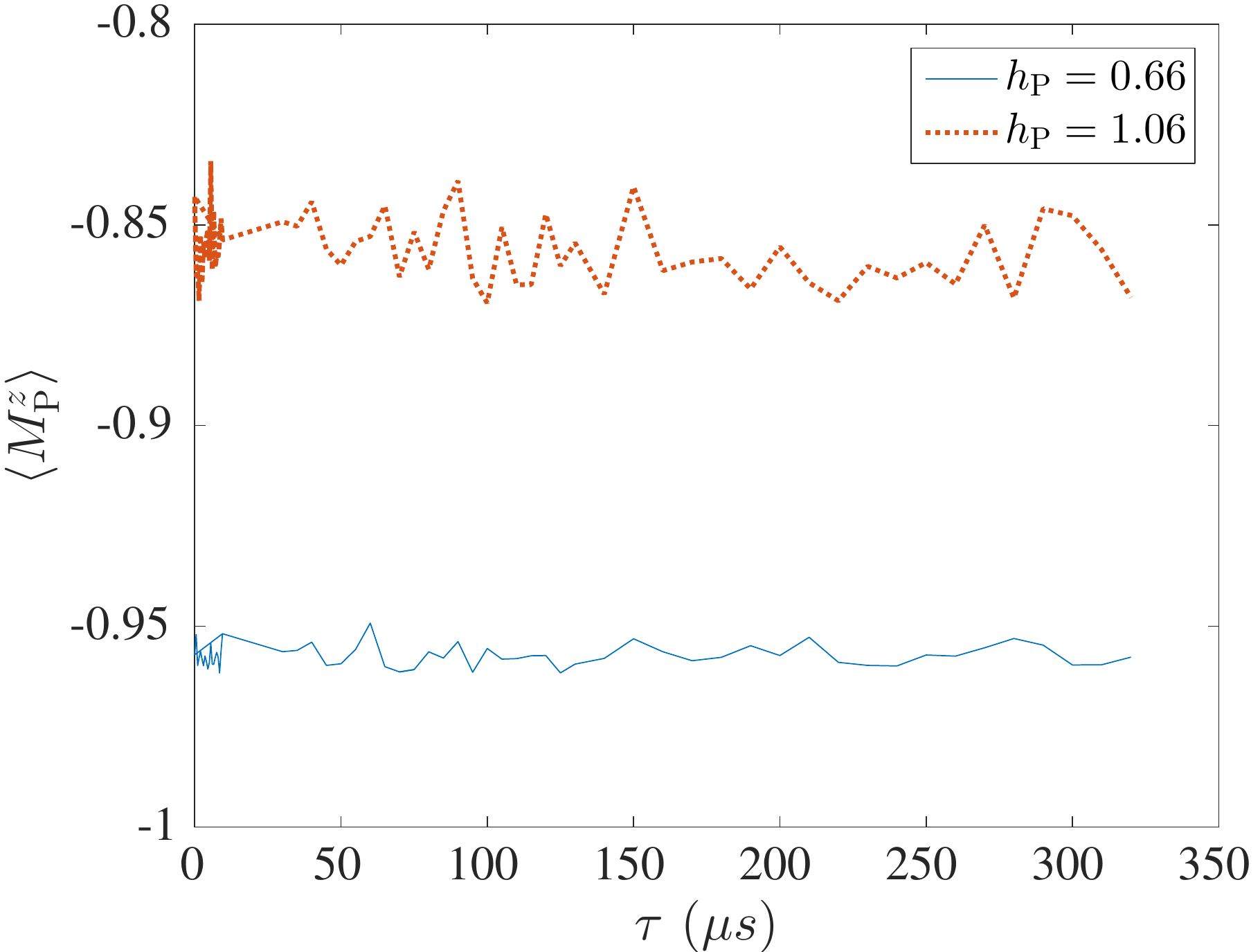}   \label{fig:O3time}}
   \subfigure[]{\includegraphics[width=0.85\columnwidth]{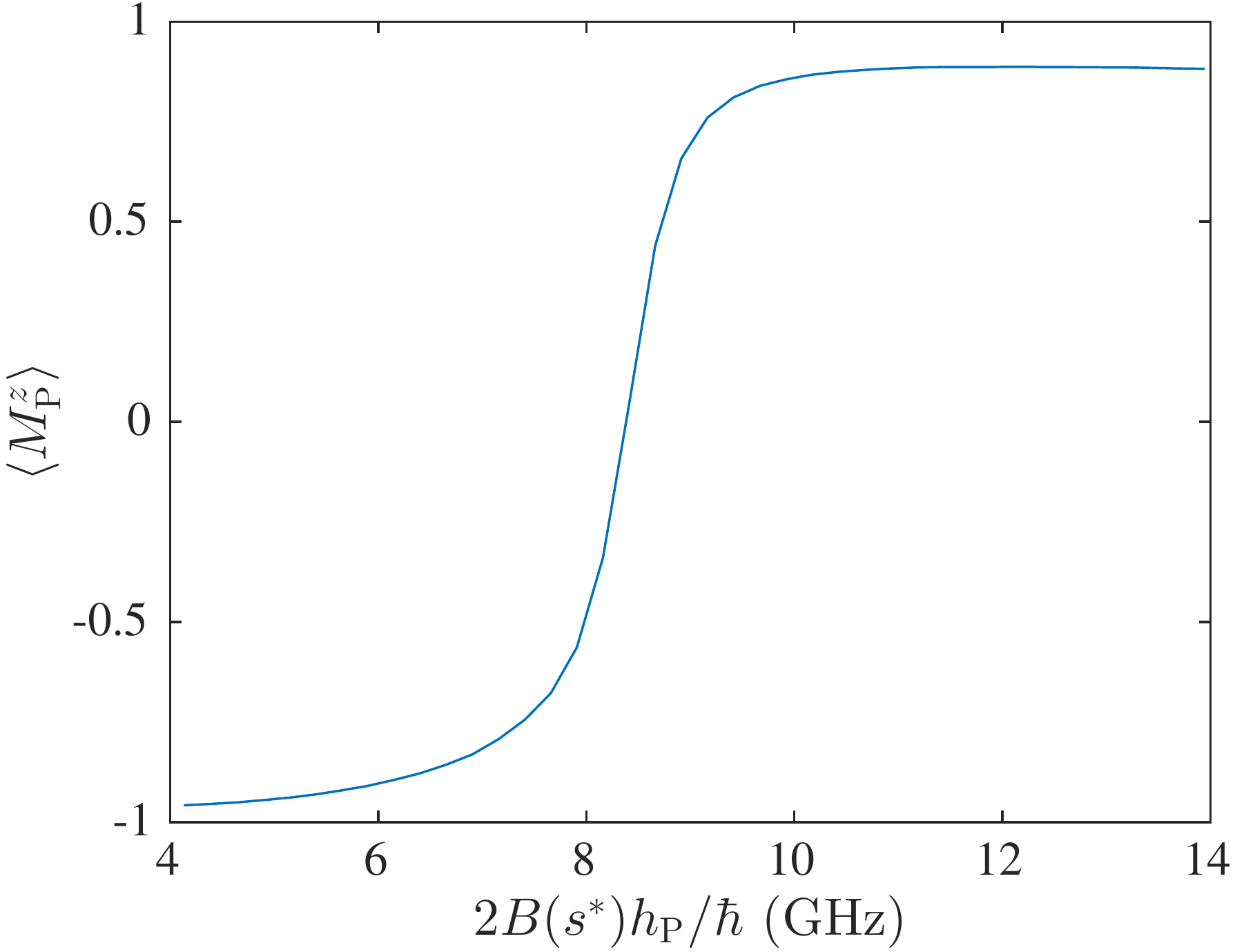}   \label{fig:O3hp}}
     \caption{(Color online) Results for the spin Langevin model described in Eq.~\eqref{eq:spin-Langevin} for the 3 qubit problem.  Simulation parameters are $\chi = 10^{-3}$ and $T = 12.5$ mK.  1000 runs were performed at each $h_\mathrm{P}$ value.  $\langle M_{\mathrm{P}}^z \rangle$ denotes the average over these 1000 runs for the $z$-component of the magnetization of the penalty qubit.}
 \end{figure*}

Since we were unable to reproduce the observed broadening we must conclude that our noise model is lacking a relevant feature, likely related to the breakdown of the weak coupling limit. Alternative methods (such as the non-interacting blip approximation \cite{Boixo:2014yu}, though it is restricted to two-level systems) may thus need to be employed or developed to capture the broadening aspect of the experiment.

\section{Spin-Langevin Model} \label{app:O3}
Given the importance of the unitary dynamics component in reproducing the experimental results, we consider an alternative classical model with dynamics, specifically a (Markovian) spin-Langevin equation \cite{Jayannavar:1990,PTPS.46.210} with a Landau-Lifshitz friction term \cite{LL:1935,PTPS.46.210},
\beq
 \frac{d}{dt} \vec{M}_i  = -\left( \vec{H}_i + \vec{\xi}(t) + \chi  \vec{H}_i \times \vec{M}_i  \right) \times \vec{M}_i \ ,
 \label{eq:spin-Langevin}
\eeq
with the Gaussian noise $\vec{\xi} = \{\xi_i\}$ satisfying
\beq
\langle \xi_i(t) \rangle = 0 \ , \quad \langle \xi_i(t)  \xi_j(t') \rangle = 2 k_B T \chi \delta_{ij} \delta(t- t') \ ,
\eeq
and 
\beq
\vec{H}_i = 2 A_i(t) \hat{x} -2 B(t) \left(h_i  + \sum_{j \neq i} J_{ij}  \vec{M}_j \cdot \hat{z}  \right) \hat{z}  \ ,
\eeq
where $\hat{x}$ and $\hat{z}$ are unit vectors. Like the SSSV model, this model is another classical limit that can be derived from the Keldysh path integral formalism \cite{Crowley:2015fr}. We follow the same procedure as described in Sec.~\ref{sec:experiment} of the main text.  We find that magnetization along the $z$-direction of the penalty qubit does not depend on the hold time $\tau$, as shown in Fig.~\ref{fig:O3time}.  Therefore, we do not observe the experimental signature of exponential decay of the all-down state.  Furthermore, the final outcome of the $z$-magnetization in fact depends smoothly on $h_\mathrm{P}$ as shown in Fig.~\ref{fig:O3hp}.  Therefore, although this model does include dynamics, it fails to reproduce the experimental signature.
%

 %

\end{document}